\documentclass[11pt,twoside]{article}  

\usepackage{asp2006}   

\usepackage{epsf}
\usepackage{psfig}
\usepackage{lscape}
\usepackage{graphicx}

\begin{document}
\title{Atmospheric Circulation of Hot Jupiters: A Review of Current
Understanding}
\author{Adam P. Showman}
\affil{Department of Planetary Sciences, Lunar and Planetary Laboratory,
University of Arizona, Tucson AZ 85721}
\author{Kristen Menou}
\affil{Department of Astronomy, 1328 Pupin Hall, 
Columbia University, New York, NY 10027}
\author{James Y-K. Cho}
\affil{Astronomy Unit, School of Mathematical Sciences, Queen Mary,
University of London, Mile End Road, London E1 4NS, U.K.}

\begin{abstract}
Hot Jupiters are new laboratories for the physics of giant planet
atmospheres. Subject to unusual forcing conditions, the circulation
regime on these planets may be unlike anything known in the Solar
System. Characterizing the atmospheric circulation of hot Jupiters is
necessary for reliable interpretation of the multifaceted
data currently being collected on these planets. We discuss several
fundamental concepts of atmospheric dynamics that 
are likely central to obtaining a solid understanding of these 
fascinating atmospheres.  A particular effort is made to compare 
the various modeling approaches 
employed so far to address this challenging problem. 
\end{abstract}

\section{Introduction}

An exploding body of observations promises to unveil the meteorology
of giant planets around other stars.  Because of their high
temperatures, short orbital periods, and likelihood of transiting
their stars, the hot Jupiters are yielding their secrets most easily,
and we now have constraints on radii, composition, albedo, dayside
temperature structure, and even day-night temperature distributions
for a variety of planets \citep[e.g.,][] {knutson-etal-2007b,
cowan-etal-2007, harrington-etal-2006, harrington-etal-2007,
charbonneau-etal-2002, charbonneau-etal-2005, charbonneau-etal-2007,
deming-etal-2005, deming-etal-2006}. 
A knowledge of atmospheric dynamics is 
crucial for interpreting these observations.

Understanding the atmospheric circulation of {\it any} planet is a
difficult task, and hot Jupiters are no exception.  The difficulty
results from the nonlinearity of fluid motion and from the complex
interaction between radiation, fluid flow, and cloud microphysics.
Turbulence, convection, atmospheric waves, vortices, and jet streams
can all interact in a complex manner across a range of temporal and
spatial scales.  Furthermore, radiative transfer and dynamics are
coupled and cannot be understood in isolation.  For example, the
equator-to-pole temperature contrasts on terrestrial planets depend
not only on the rate of latitudinal energy transport by the
circulation but also on the way the atmospheric radiation
field is affected by the advected temperature field.  Cloud
microphysics complicates the problem even more, because the
large-scale radiative properties of clouds depend on the microphysics
of the cloud particles (particle number density, shape, size
distribution, and absorption/scattering properties).  We thus have a
non-linear, coupled radiation-hydrodynamics problem
potentially involving interactions over 14 orders of magnitude, from
the cloud-particle scale ($\sim1\,\mu$m) to the global scale
($\sim10^8\,$m for hot Jupiters).

It is worth emphasizing here the important difference between
the complexity of planetary atmospheres and the relative simplicity
of stars, exemplified by the main sequence in the HR diagram.  While
specifying a few global parameters---such as mass, composition, 
and age---is typically sufficient to understand the key observable
properties of stars, this is generally not the case for planetary
atmospheres. It is possible that a significant observable diversity
exists even among a group of extrasolar planets which share similar
global attributes.  Understanding such complexity is a challenge for
extrasolar planetary science.

A proven method for dealing with the complexity of
planetary atmospheres is the concept of a {\it model hierarchy.}
Even a hypothetical computer model that included all relevant
processes and perfectly simulated an atmosphere would not, by itself,
guarantee an {\it understanding} of the atmospheric behavior any more
than would the observations of the real atmosphere \citep{held-2005}.
This is because, in complex numerical simulations, it is
often unclear how and why the simulation produces a specific behavior.
To discern which processes cause which outcomes, it is important to compare
models with a range of complexities (in which various processes
are turned on or off) to build a hierarchical understanding.

For example, east-west jet streams occur in the atmospheres of 
all the planets in our Solar System.  The study of these jets
involves a wide range of models, all of which have important
lessons to teach.  The most idealized are pure 2D models 
that  investigate jet formation in the simplest 
possible context of horizontal, 2D turbulence interacting with 
planetary rotation \citep[e.g.,][]{williams-1978, yoden-yamada-1993,
cho-polvani-1996b, huang-robinson-1998,
sukoriansky-etal-2007}.    Despite the fact that
these simplified models exclude vertical structure, thermodynamics, 
radiation, and clouds and provide only a crude parameterization of 
turbulent stirring, they produce jets with several similarities to those
on the planets.  Next are 
one-layer ``shallow-water-type'' models that allow
the fluid thickness to vary, which introduces
buoyancy waves, alters the vortex interaction lengths, and hence
changes the details of jet formation 
\citep{cho-polvani-1996a, cho-polvani-1996b, scott-polvani-2007,
showman-2007}.  3D models with simplified forcing allow the investigation 
of jet vertical structure, the interaction of heat transport and jet formation,
and the 3D stability of jets to various instabilities
\citep{cho-etal-2001, williams-1979, williams-2003a, schneider-2006, 
lian-showman-2007}.  
Such models suggest, for example, that Jupiter
and Saturn's superrotating\footnote{Superrotation is defined simply
as a positive (eastward) longitudinally averaged wind velocity at the equator,
so that the atmospheric gas rotates faster than the planet.} 
equatorial jets may require 3D dynamics.
Finally are full 3D general-circulation models (GCMs) that include
realistic radiative transfer, representations of clouds, and other effects
necessary for detailed predictions and comparisons with observations
of specific planets. The comparison of simple models with
more complex models provides  insights not easily obtainable 
from one type of model alone.  This lesson applies equally
to hot Jupiters: a hierarchy of models ranging from simple
to complex will be necessary to build a robust understanding.

Here we review our current understanding of atmospheric circulation
on hot Jupiters.  We first describe basic aspects of relevant theory from
atmospheric dynamics; this is followed by a detailed comparison of
the equations, forcing methods, and results obtained by the different
groups attempting to model the atmospheric circulation on these fascinating
objects.

\section{Basic considerations}

Despite the complexity of atmospheric circulation, there exists 
a 40-year history of work in atmospheric dynamics
on solar-system planets that can guide our 
understanding of hot Jupiters.

\medskip

\noindent\underline{\it Differences between hot Jupiters and
solar-system giants:} Jupiter, Saturn, Uranus, and Neptune rotate
rapidly, with rotation periods ranging from $10\,$hours (Jupiter)
to $17\,$hours (Uranus). This implies rotationally dominated flows.
In contrast, hot Jupiters are expected to rotate synchronously with 
their orbital periods \citep{guillot-etal-1996}, at least when
orbital eccentricity is small, implying rotation periods of 
1--$5\,$days for the known transiting planets.  Some hot Jupiters
will therefore be rotationally dominated, while on others 
the Coriolis forces will have more modest importance.  Note,
however, that even ``slowly'' rotating hot Jupiters rotate much
faster than Titan and Venus (rotation periods of 16 and 243 Earth days, 
respectively).  On most hot Jupiters, rotation will have
a major impact in modifying the flow geometry.

The forcing on hot Jupiters differs greatly from that on solar-system
giants. First is simply the amplitude of the radiative forcing:
HD209458b absorbs and reradiates almost $\sim260,000\,{\rm W}\,
{\rm m}^{-2}$ on a global average, as compared to $240\,{\rm W}\,
{\rm m}^{-2}$ for Earth, $14\,{\rm W}\,{\rm m}^{-2}$ for Jupiter,
and $0.7\,{\rm W}\,{\rm m}^{-2}$ for Uranus and Neptune.
A second difference is that, on the giant planets in our 
Solar System, the flow seems to be driven by 
convection, thunderstorms, or baroclinic instabilities with small
length scales (see below).  Jupiter, Saturn, and
Neptune have comparable instrinsic and total heat fluxes, allowing
interior convection to greatly affect the atmosphere; the convective
zone extends to the visible cloud layers at $\sim1\,$bar on these 
planets. For hot
Jupiters, however, a statically stable radiative zone extends to
pressures of 100--$1000\,$bars, separating the observable atmosphere
from the convective interior; furthermore, the interior convective
forcing is $10^4\,$ times weaker than the stellar insolation.  Thus,
the hemispheric-scale stellar forcing is likely to be more dominant on
hot Jupiters than for the giant planets in our Solar
System. Finally, the expected radiative times above the
photospheres on hot Jupiters are extremely short --- days or less ---
whereas those on solar-system giants are years to decades.
All these differences will impact the flow in ways that remain
to be understood.

Despite our ability to list these differences in forcing,
the way the circulation responds to these differences
is often subtle.  The planets in our Solar System demonstrate
that the relationship between forcing and response is nontrivial.
For example, Neptune and Uranus exhibit similar wind patterns
but their forcings differ greatly.  Neptune's intrinsic heat flux
is 1.6 times the absorbed sunlight, implying a strong role for interior 
convection; on Uranus, in contrast, the intrinsic flux is $<10\%$
of the total flux, implying that external forcing is predominant.
Moreover, the small ($30^{\circ}$) obliquity on Neptune implies
the existence of a regular day-night pattern where the equator
receives more sunlight than the poles; in contrast, the
$98^{\circ}$ obliquity on Uranus means that the poles 
receive greater sunlight than the equator, and that much of
the planet experiences decades of near-continuous 
sunlight followed by decades of darkness. 
This Uranus-Neptune comparison suggests that a rotating-stratified atmospheric 
flow can respond to forcing in a manner which is not simply dictated by
the forcing geometry.

\medskip

\noindent
\underline{\it Flow length scales:} Theoretical work has demonstrated
the importance of two fundamental length scales in atmospheric
dynamics.  First, the {\it Rhines length}, defined as
$(U/\beta)^{1/2}$, is the scale at which planetary rotation causes
east-west elongation (jets).  Here $U$ is a characteristic horizontal
wind speed and $\beta\equiv2\Omega\cos\phi/a$ is the latitudinal
gradient of planetary rotation ($\Omega$ is the planetary angular
rotation rate, $\phi$ is the latitude, and $a$ is
the planetary radius).  Turbulence at small scales tends to
be horizontally isotropic, but at length scales approaching the Rhines scale,
turbulent structures grow preferentially in the east-west direction
rather than the north-south direction.  This typically leads to a flow
exhibiting a banded structure with east-west jet streams.

Second, the {\it Rossby deformation radius}, defined as $NH/f$,
is the scale at which pressure perturbations are resisted by the
Coriolis forces ($f\equiv2\Omega\sin\phi$ is the Coriolis parameter,
$H$ is the scale height, and $N$ is the 
Brunt-V\"ais\"al\"a
frequency, i.e., the oscillation frequency for a vertically displaced 
air parcel in a statically stable atmosphere).  Vortices
often have horizontal sizes near the deformation radius.  And
baroclinic instabilities --- a form of sloping convection
that can occur in the presence of horizontal temperature contrasts,
converting potential energy to
kinetic energy --- typically generate turbulent eddies 
with a horizontal size comparable to the deformation radius.  
Such instabilities 
are the primary cause of midlatitude weather on Earth and 
Mars and may be important in driving the jet streams on Jupiter, 
Saturn, Uranus, and Neptune \citep{williams-1979, williams-2003a, 
friedson-ingersoll-1987, lian-showman-2007}.

On Jupiter and Saturn, the deformation radius and Rhines length are,
respectively, $\sim2000$ and $\sim10,000\,$km 
(i.e., much less than the planetary radius). This explains
why these planets have narrow jets, numerous cloud bands, 
and small vortices.  On most planets, including Earth's atmosphere,
Earth's oceans, and the atmospheres on Mars, Jupiter, and Saturn, 
there exists a large population of flow structures with sizes near the
deformation radius.

In contrast, most hot Jupiters have modest rotation rates, 
large scale heights (up to ten times that on solar-system giants 
as a result of the high temperature), and strong vertical 
stratification (a result of the dominant external irradiation).
This means that, on most hot Jupiters, the 
Rhines length and Rossby deformation radius are close to the planetary 
radius.  Unlike Jupiter and Saturn, the dominant flow 
structures on hot Jupiters should therefore be planetary in scale 
\citep{showman-guillot-2002, menou-etal-2003}.

The length scales discussed here will strongly influence the atmospheric
energetics.  In a 3D turbulent fluid, vortex stretching
occurs readily, generating small-scale flow structures from 
large-scale ones and
causing energy to cascade downscale.  In a quasi-2D fluid such as an
atmosphere, however, vortex stretching is inhibited, and energy
instead experiences an {\it inverse cascade} that transfers energy from
small to large length scales \citep{pedlosky-1987,vallis-2006}.  
(The simplest example is vortex merger, which gradually produces larger
and larger vortices over time.) The extent to which the inverse
cascade occurs depends on the length scale of energy injection relative
to the planetary size: if energy injection occurs at small scales (as
may be the case on Jupiter and Saturn), the inverse cascade
occurs readily.  On the other hand, if energy injection occurs at
planetary scales (as may occur on hot Jupiters), the inverse cascade
could be less dominant (but see \S3.2 below).

\medskip

\noindent
\underline{\it Wind speeds:} Despite progress, there is still no
general theory for what sets the wind speeds on the giant planets in
our Solar System.  The speeds on all four giants exceed those on
Earth, which presumably relates to the lack of a surface
(and its associated friction) on the giant planets.  More puzzling is
the observation that the wind speeds on Neptune exceed those on Jupiter
despite the fact that Neptune's total atmospheric heat flux ---
which ultimately drives the circulation --- is only 4\% of
that on Jupiter.  A lesson is that the mean speed involves an
equilibrium between {\it forcing} and {\it damping} (e.g., friction).
Fast winds can occur with weak forcing if the damping is also weak;
alternatively, slow winds can occur in the presence of strong forcing 
if the damping is strong.  The damping mechanisms on solar-system 
giants (which include turbulent mixing and radiative cooling to space) 
seem to be relatively weak, leading to fast winds.
However,  it is not fully understood how these mechanisms translate 
into the hot Jupiter context, particularly since the strength of
the damping (as well as the forcing) depends nonlinearly on the 
flow itself.  Careful theoretical work can shed light on how
damping and forcing interact to produce an equilibrated flow
speed relevant for hot Jupiters.

Despite the above caveats, the known dynamical link between winds
and temperatures can be invoked to make useful statements about
the winds.  
In rotating atmospheres, horizontal temperature contrasts are linked to
vertical differences in horizontal wind via the {\it thermal-wind relation}
\citep{holton-2004}. (This is often defined in the context of geostrophic
balance but it can be extended to a more general balance.) 
On terrestrial planets, the speed is zero 
at the surface.  So, given an estimate of the equator-to-pole temperature
difference, the thermal-wind relation can provide a crude
estimate of the actual wind speeds.  The forcing and damping
processes are relatively well understood for terrestrial planets,
and fully nonlinear 3D circulation models do a reasonably good job
of simulating the structure and speeds of the winds.  


For giant planets, one can profitably apply the thermal-wind relation
to obtain the difference in horizontal wind speed between (say) the
cloud deck and the base of the weather layer (thought to be at
$\sim10\,$ bars for Jupiter) \citep[e.g.,][]{ingersoll-cuzzi-1969}.
The difficulty on giant planets is that, unlike the terrestrial
planets, the speed at the base of this layer is not necessarily zero
--- the planet may exhibit a so-called
``barotropic\footnote{A barotropic fluid is one where the
contours of constant pressure and density align, while a 
baroclinic fluid is one where these contours misalign.  The
convective interior has nearly constant entropy and is hence
barotropic. Under sufficiently rapid rotation this leads to the
Taylor-Proudman theorem in the convective interior, 
which states that east-west wind is constant in columns that
penetrate the molecular envelope
in the direction parallel to the rotation axis.  This is called
a barotropic wind.}''  
wind that penetrates through the convective
molecular envelope.  In this case, the total wind at the cloud tops
would be the sum of the thermal-wind (or ``baroclinic'' component) and
the barotropic wind.  

In the context of hot Jupiters, several numerical simulations have been
performed that are driven by the intense dayside heating and nightside
cooling \citep{showman-guillot-2002, cooper-showman-2005,
cooper-showman-2006, dobbs-dixon-lin-2007, langton-laughlin-2007}.
These simulations are effectively constraining the thermal-wind
component of the flow (the height-variable component that exists in
the radiative zone above $\sim100$--$1000\,$bars). All of these
simulations obtain wind speeds of $\sim1\,{\rm km}\,{\rm sec}^{-1}$,
in agreement with order-of-magnitude estimates using the thermal-wind
relation \citep{showman-guillot-2002}.  All of the published
simulations assume that the deep barotropic wind component is zero.  
It is worth bearing in mind, however, that the barotropic wind component is
unknown, and if it is strong then the actual wind speed 
could differ in magnitude and geometry from that suggested by these 
simulations.  Detailed comparisons of simulations with observed
lightcurves and spectra of hot Jupiters can provide constraints
on the extent to which the models are capturing the correct flow
regime.

\medskip

\noindent
\underline{\it Temperatures:} Insight into temperature patterns
may be gained with timescale arguments.  Suppose $\tau_{\rm
rad}$ is the radiative timescale (i.e., the time for radiation
to induce large changes in the entropy) and $\tau_{\rm advect}$ is the
timescale for air to advect across a hemisphere (e.g., time for air to
travel from dayside to nightside or equator to pole).  If $\tau_{\rm
rad} \ll \tau_{\rm advect}$, then large temperature contrasts are
expected, whereas if $\tau_{\rm rad} \gg\tau_{\rm advect}$, then
temperatures should be horizontally homogenized.

To illustrate the usefulness of these arguments,
Table~1 lists these timescales and temperature contrasts for 
the tropospheres of the terrestrial planets.  
The numbers show general agreement
with the above arguments.  The radiative time constants 
are long on Venus, intermediate on Earth, and short on Mars.
Regarding transport, we must distinguish longitude from latitude.
On Earth and Mars, the planetary rotation dominates the transport 
of air parcels in longitude, so parcels cycle between day and night 
with a timescale of $1\,$day.   Venus rotates slowly, but winds and 
rotation nevertheless transport air from dayside to nightside over 
a period of days to months.  The numbers in the Table imply that
$\tau_{\rm rad}\gg \tau_{\rm advect,lon}$
for Venus and Earth; in agreement, longitudinal temperature
variations are small on these two planets.  (They reach $\sim10\,$K
on Earth, but this results largely from the uneven continent-ocean
distribution; a continent-free ``aquaplanet'' Earth would exhibit 
even smaller longitudinal temperature variations.)  Latitudinal transport
timescales (Table 1) imply
that $\tau_{\rm rad}> \tau_{\rm advect,lat}$ for Venus,
$\tau_{\rm rad}\sim \tau_{\rm advect,lat}$ for Earth, and
$\tau_{\rm rad}< \tau_{\rm advect,lat}$ for Mars.  In qualitative
agreement with expectations, equator-to-pole temperature contrasts 
are small on Venus, intermediate on Earth, and large on Mars
(Table 1).

\begin{table}[!ht]
\caption{Timescales and temperature contrasts on terrestrial planets}
\smallskip
\begin{center}
{\small
\begin{tabular}{cccccc}
\tableline
\noalign{\smallskip}
Planet & $\tau_{\rm rad}$ & $\tau_{\rm advect,lon}$ & $\Delta T_{\rm lon}$ &$\tau_{\rm advect,lat}$ &$\Delta T_{\rm lat}$\\
\noalign{\smallskip}
\tableline
\noalign{\smallskip}
Venus & years & days & $<$few$\,$K &months? &few$\,$K\\
Earth & weeks--months  &1 day & $\sim10\,$K &weeks &20--40$\,$K\\
Mars & days  &1 day &$\sim20$--$40\,$K &days--weeks &$\sim50$--$80\,$K\\
\noalign{\smallskip}
\tableline
\end{tabular}\\
Values refer to tropospheres (0.005 bars for Mars, 1 bar for
Earth, and $\sim10\,$bars on Venus). In this table, ``day''
refers to Earth day ($86400\,$sec). Refs: \citet{gierasch-etal-1997, 
peixoto-oort-1992, read-lewis-2004}.  
}
\end{center}

\end{table}

What about giant planets?  The four solar-system giants 
exhibit latitudinal temperature contrasts above the clouds
of only $\sim5\,$K 
\citep{ingersoll-1990}.  This results from
several processes.  First, in the interiors, convective
mixing times are short ($\sim$decades) but the radiative timescales
are geological, so the interior entropy is homogenized
\citep{ingersoll-porco-1978}.  On Jupiter, Saturn, and Neptune 
(which have internal convective heat fluxes comparable to the
absorbed solar flux), the top of the 
constant-entropy interior is close to the photosphere; thus,we
expect nearly constant photospheric temperatures across the planet.
Second, on all four giants, the radiative timescales above the clouds 
are long (years to decades), and so even relatively sluggish 
circulations are capable of homogenizing the temperature patterns.  
The latter mechanism may be particularly important on Uranus,
which lacks a significant interior heat flux.

Hot Jupiters differ from solar-system giants because the
radiative timescales above the photosphere are short and because
the homogenized interior is hidden below a radiative zone extending to
$\sim100$--$1000\,$bars \citep{guillot-etal-1996,
guillot-showman-2002, bodenheimer-etal-2001, bodenheimer-etal-2003,
baraffe-etal-2003, chabrier-etal-2004, iro-etal-2005}, which is far
below the expected photosphere pressure of
$\sim0.01$--1$\,$bar.  Large day-night temperature differences are 
therefore possible at the photosphere depending on $\tau_{\rm rad}$ and
$\tau_{\rm advect}$ there.  Current estimates, which derive from
radiative calculations that neglect circulation, suggest that
$\tau_{\rm rad}$ is extremely pressure dependent, ranging from
$\sim1\,$hour at $10^{-3}\,$bars to $\sim1\,$year at $10\,$bars
\citep{iro-etal-2005}.  Considering wind speeds ranging from
0.3--$3\,{\rm km}\,{\rm sec}^{-1}$ then implies that $\tau_{\rm
rad}\sim\tau_{\rm advect}$ at pressures of $\sim0.1$--$1\,$bar.  Above
this level, day-night temperature differences should be large, while
below this level the temperature differences could become small
\citep{showman-guillot-2002}.

A major caveat, of course, is that one must have knowledge of the
circulation to estimate $\tau_{\rm advect}$.  Even $\tau_{\rm rad}$
can depend on the circulation through its dependence on temperature
and composition (which in turn depend on the circulation).  In
the absence of other information, the above arguments are thus
generally insufficient to make rigorous predictions of temperature
contrasts on planets, since independent timescales are assigned to
radiation and hydrodynamics while the two processes are in fact
intimately coupled, as mentioned earlier.  Nevertheless, the
arguments can be used {\it a posteriori} to understand the temperature
patterns occurring in observations --- or in detailed nonlinear
numerical simulations of the circulation --- when these observations
and/or simulations are sufficient to define $\tau_{\rm advect}$ and
$\tau_{\rm rad}$.

\section{Models of atmospheric circulation on hot Jupiters: Similarities
and differences}

A variety of approaches have been used to study the atmospheric
circulation on hot Jupiters.  These include solutions of the global
two-dimensional shallow-water equations on a rotating sphere
\citep{langton-laughlin-2007}, the global three-dimensional primitive
equations\footnote{The leading-order dynamical equations
solved in meteorology and climate studies are called the primitive
equations \citep[e.g.,][]{holton-2004}; see \S3.1 and \S3.5.} 
on a rotating sphere
\citep{showman-guillot-2002, cooper-showman-2005,cooper-showman-2006},
the global equivalent-barotropic formulation of these primitive
equations on the rotating sphere \citep{cho-etal-2003, cho-etal-2007},
the two-dimensional Navier-Stokes equations neglecting rotation in an
equatorial cross section \citep{burkert-etal-2005}, and the
three-dimensional Navier-Stokes equations omitting the polar regions
of a rotating sphere \citep{dobbs-dixon-lin-2007}.  The forcing in all
of these models is simplified; no models yet include realistic
radiative transfer.

\subsection{Work of Showman, Cooper and collaborators}
\citet{showman-guillot-2002}, \citet{cooper-showman-2005}, and
\citet{cooper-showman-2006} solved the 3D primitive equations on the
whole sphere for parameters relevant to HD209458b. The 3D
approach is motivated by the expectation that the radiative
time constant varies by orders of magnitude in the vertical
\citep{iro-etal-2005}, which in the presence of the day-night
heating gradient would cause patterns of temperature
and wind that vary both vertically and horizontally in an 
inherently 3D manner.  

The primitive
equations are the relevant equations for large-scale flow in
statically stable atmospheres whose horizontal dimensions greatly
exceed the vertical dimensions.  This is expected to be true on hot
Jupiters, which have horizontal scales of $10^7$--$10^8\,$m but
atmospheric scale heights of only 200--$500\,$km (depending on
temperature and gravity), leading to a horizontal:vertical aspect
ratio of 20--500.  This large aspect ratio allows the vertical
momentum equation to be replaced with local hydrostatic balance,
meaning that the {\it local} vertical pressure gradient $\partial
p/\partial z$ is balanced by the local fluid weight $\rho g$, where
$p$, $\rho$, $g$, and $z$ are pressure, density, gravity, and height
\citep[see for example the derivation in][and \S3.5 below]{holton-2004}.  
Lateral and height variations in these
quantities can still occur and evolve dynamically with the flow.  The
primitive equations admit the full range of balanced motions, gravity
(i.e. buoyancy) waves, rotationally modified (e.g., Rossby and Kelvin)
waves, and horizontally propagating sound waves, but they filter out
vertically propagating sound waves.

\citet{showman-guillot-2002}, \citet{cooper-showman-2005}, and
\citet{cooper-showman-2006} forced the flow using a Newtonian
heating/cooling scheme, which parameterizes the thermodynamic heating
rate as $(T_{\rm eq}-T)/\tau_{\rm rad}$, where $T_{\rm eq}$ is the
specified radiative-equilibrium temperature profile (hot on the
dayside, cold on the nightside), $T$ is the actual temperature, and
$\tau_{\rm rad}$ is the radiative-equilibrium timescale (a function of
pressure).  The vertical structure of $T_{\rm eq}$ and $\tau_{\rm
rad}$ were taken from \citet{iro-etal-2005}; the day-night difference
$T_{\rm eq}$ was a free parameter that was varied from 100 to
$1000\,$K.  This formulation, of course, does not represent a rigorous
treatment of radiative transfer, but it provides a physically
motivated means to investigate how the expected day-night thermal
forcing would drive the atmospheric circulation.

\begin{figure}[htb]
\begin{center}
\includegraphics[angle=-90, scale=0.5]{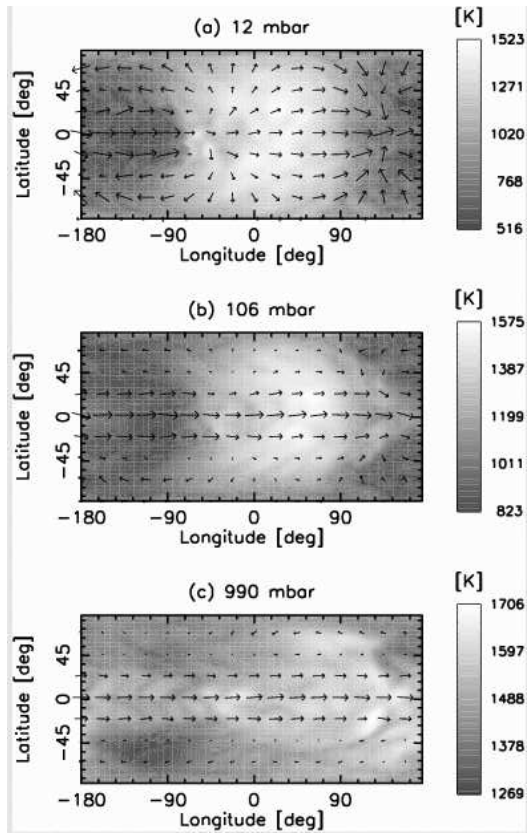}
\caption{Simulated temperature (greyscale) and horizontal winds 
(arrows) for three pressure levels 
from \citet{cooper-showman-2006}.  Planetary parameters of
HD209458b were adopted.}
\end{center}
\end{figure}

Figure~1 shows the temperature (greyscale) and winds (arrows) for
three layers (10, 100, and 1000 mbar from top to bottom, respectively)
after a simulated time of 1000 Earth days for a case with an
assumed day-night difference in $T_{\rm eq}$ of $1000\,$K.  
The imposed heating
contrast leads to winds of several km$\,{\rm sec}^{-1}$, and the
day-night temperature contrasts reach $500\,$K or more.  The
flow structure has a strongly 3D character.  At the top
(pressures less than $\sim100\,$mbar), the radiative time constant is
shorter than the time for winds to advect the flow, and the
temperature pattern tracks the heating --- hot on the dayside and cold
on the nightside --- despite the fast winds.  At greater pressure,
however, the radiative time constant is longer, and dynamics blows the
hottest region of the atmosphere downwind from the substellar point by
20--$60^{\circ}$ of longitude depending on altitude.  In agreement
with the Rhines-length and deformation-radius arguments presented in
\S2, the simulations exhibit only a small number of broad jets.
Interestingly, the development of a superrotating (eastward)
equatorial jet is a robust feature in all the simulations of
\citet{showman-guillot-2002}, \citet{cooper-showman-2005}, and
\citet{cooper-showman-2006}.  The mechanism for driving the jet
seems to involve equatorward pumping of eddy momentum that occurs in
conjunction with the longitudinally varying heating. This process has
previously been observed in simplified two-layer models from the Earth
sciences \citep{suarez-duffy-1992, saravanan-1993}.

\begin{figure}
\plottwo{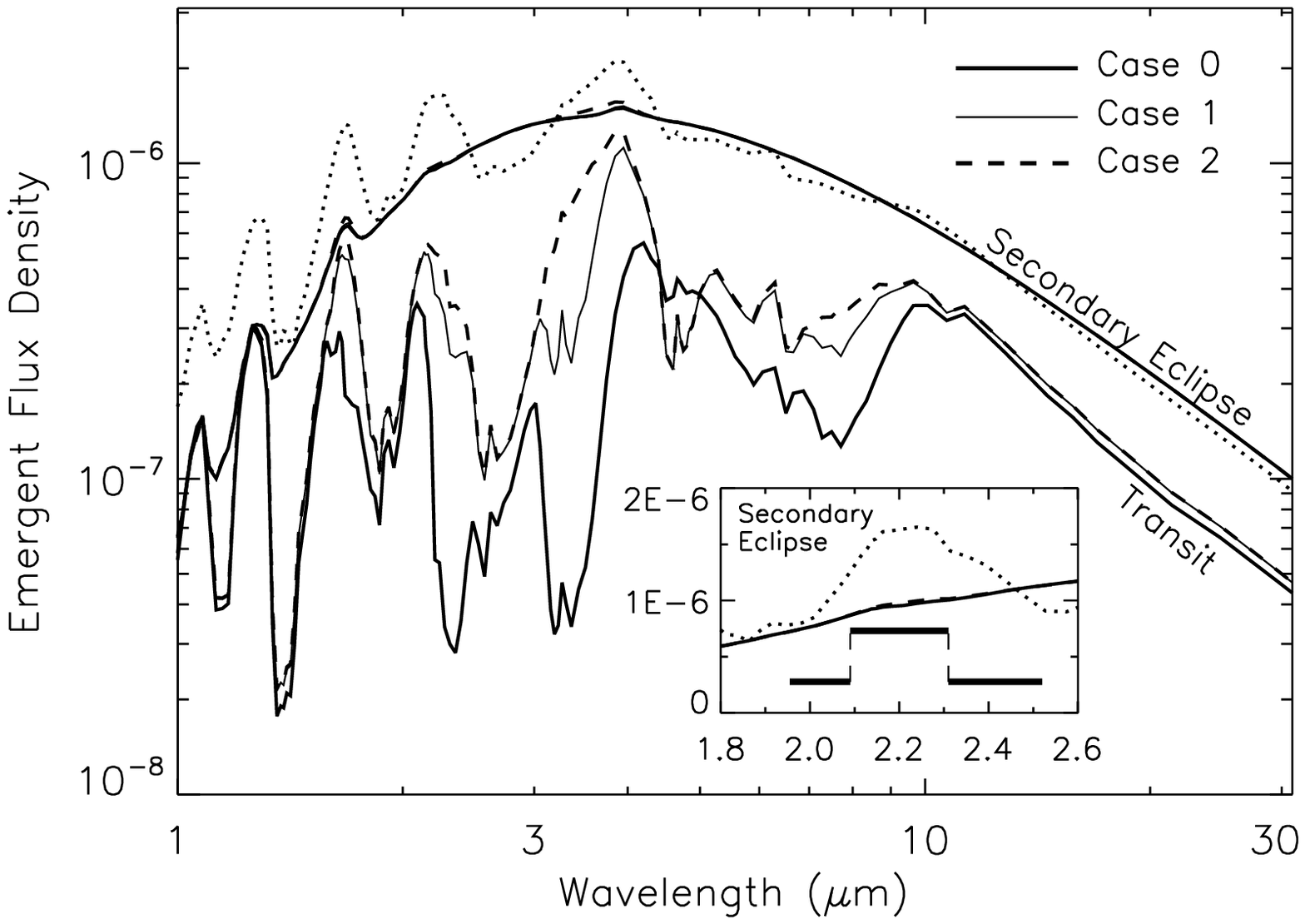}{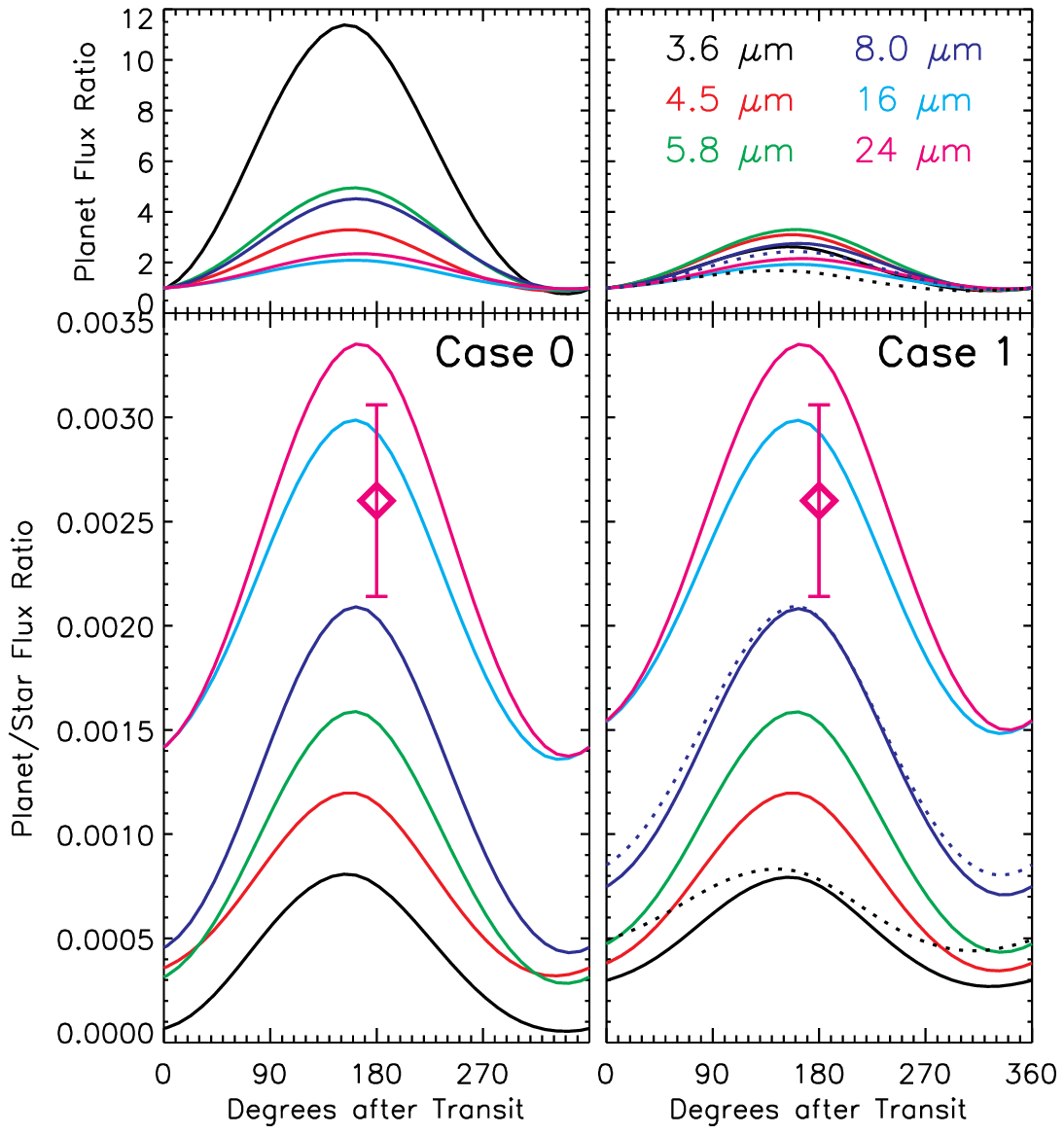}
\caption{Theoretical infrared spectra (left) and lightcurves in Spitzer
IRAC and MIPS bands (right) as calculated by \citet{fortney-etal-2006}
from the 3D simulations of \citet{cooper-showman-2006} for HD209458b. 
The dayside and nightside spectra (labelled ``secondary eclipse'' and
``transit'', respectively) differ greatly due to
the distinct vertical temperature-pressure structures in each region. 
Case 0, 1, and 2 refer to different 
assumptions about the chemistry.  In the lightcurves, note the
phase shift of the peak flux from the time near secondary eclipse
(at $180^{\circ}$).  See \citet{fortney-etal-2006}.}
\end{figure}

By pushing the temperature pattern far from radiative equilibrium, the
circulation has major implications for the infrared lightcurves and
spectra. Using a two-stream plane-parallel radiative-transfer code,
\citet{fortney-etal-2006} calculated infrared lightcurves from the 3D
temperature patterns of \citet{cooper-showman-2005,
cooper-showman-2006}.  These calculations predict that the peak IR
emission leads the secondary eclipse by $\sim2$--$3\,$hours in most
Spitzer IRAC bands (Fig.~2).  This results from advection of
the temperature field by the dynamics, which displaces the hottest
regions to the east of the substellar point (Fig.~1).
  Intriguingly, the predicted offset in
the flux peak is similar to that in the observed lightcurve for
HD189733b, but the simulations overpredict the day-night flux
difference and fail to produce to cold region to the west of the
antistellar point \citep{knutson-etal-2007b}.  

The dynamics also
drives the vertical temperature profile far from radiative
equilibrium.  As air advects from nightside to dayside in these
simulations, the air aloft warms faster than air at depth
(because of shorter radiative time constants aloft), 
producing a quasi-isothermal region on the
dayside despite the absence of such a feature in pure
radiative-equilibrium models.  This leads to dayside and nightside
spectra (Fig.~2, left) that differ substantially from 
the predictions of 1D radiative-equilibrium models 
\citep{burrows-etal-2005, barman-etal-2005,
seager-etal-2005, fortney-etal-2005}.

\subsection{Work of Cho, Menou and collaborators}

In geophysical fluid dynamics, an important distinction is made
between flows that are fundamentally 3D and flows whose essence is
largely 2D (and hence can be captured in a 2D model), even though they
are of course occurring in a 3D atmosphere.  For example, the Earth's
midlatitude troposphere experiences baroclinic instability, which is a
fundamentally 3D process and requires a 3D model to capture.  Large
parts of the Earth's stratosphere (the region overlying the
troposphere), on the other hand, lack baroclinic instabilities
because of the strong static stability.  Many aspects of the
stratosphere---such as the dynamics of the polar vortex, Rossby wave
breaking, and
turbulent mixing---have been modeled to great advantage with 2D or
quasi-2D models \citep{juckes-mcintyre-1987, juckes-1989,
  polvani-etal-1995}.  Because of their strong static stabilities and
large Rossby deformation radii (which tend to discourage baroclinic
instabilities), \citet{cho-etal-2003, cho-etal-2007} and
\citet{menou-etal-2003} have proposed that the basic flow regime of
hot Jupiter atmospheres may likewise be successfully captured in a
quasi-2D, one-layer formulation.

\begin{figure}[!ht]
\begin{minipage}[t]{4cm}
\begin{center}
\includegraphics[width=6cm, angle=-90]{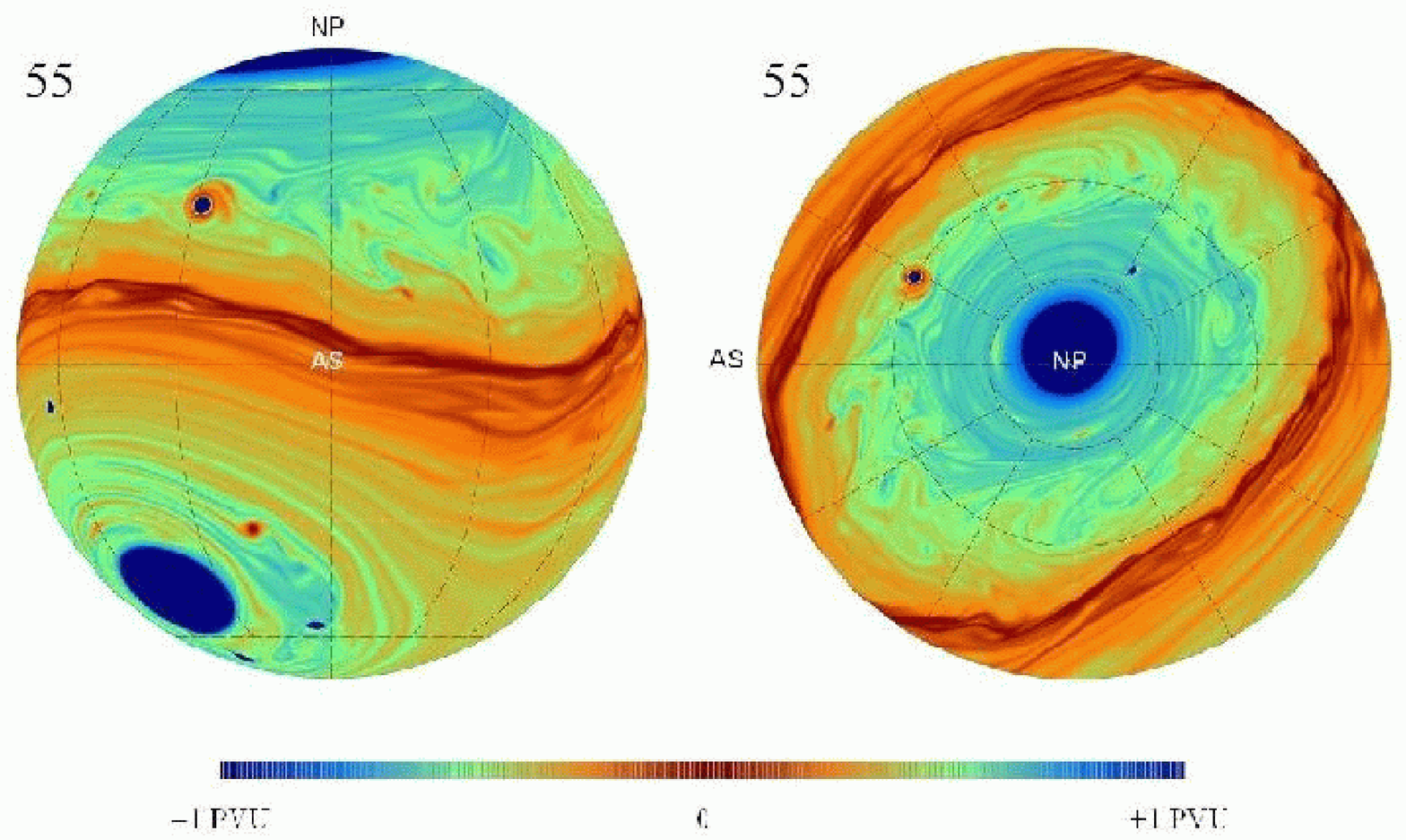}
\end{center}
\end{minipage}
\hspace{+4.5cm}
\begin{minipage}[t]{4cm}
\begin{center}
\vspace{1.3cm}

\includegraphics[width=5cm, angle=0]{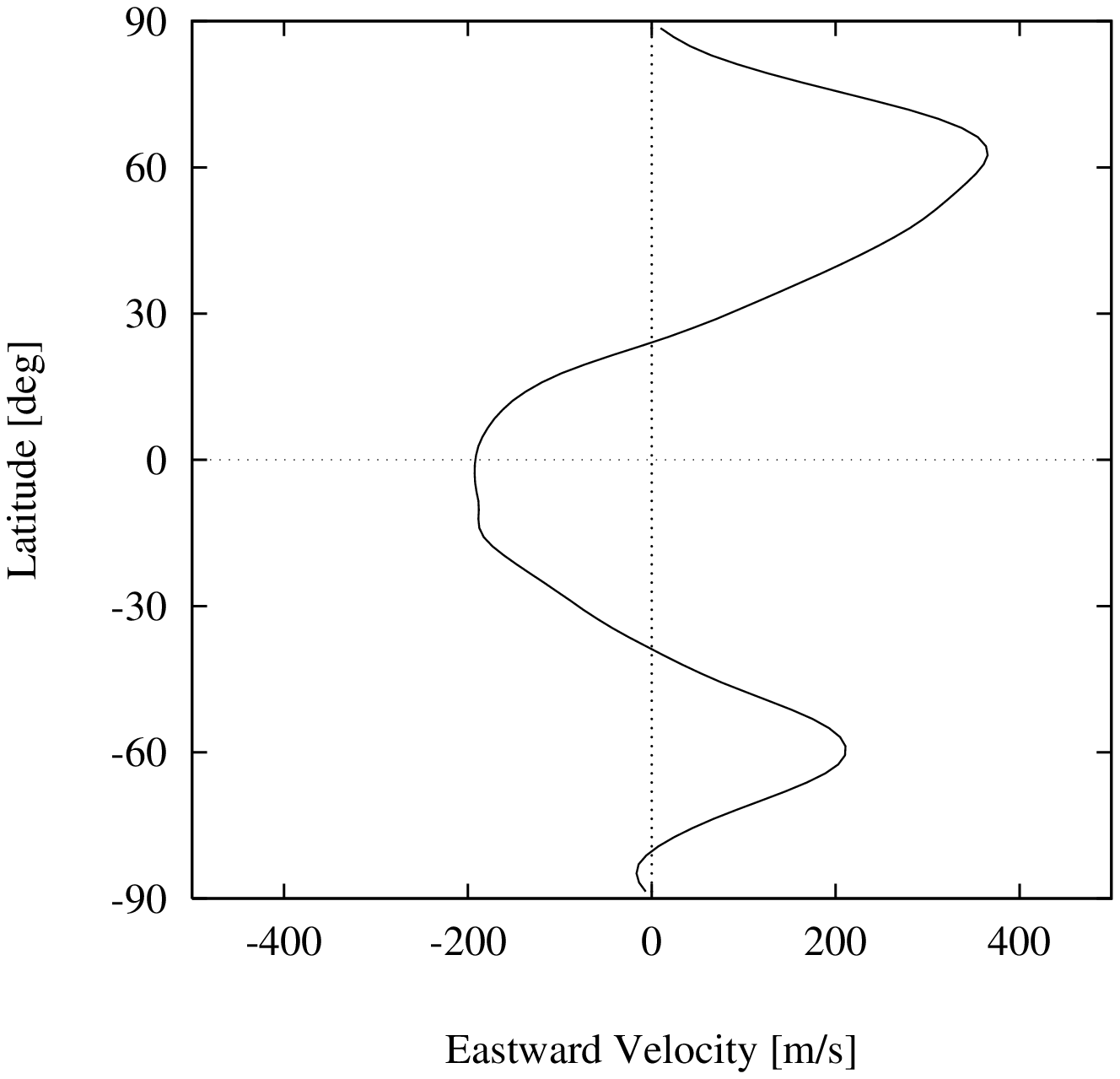}
\end{center}
\end{minipage}
\caption{[Left] Equatorial and polar views of potential vorticity (a
 flow tracer) in a specific hot Jupiter model from Cho et al. (2003,
 2007). Note the prominent circumpolar vortices formed as a result of
 potential vorticity conservation. [Right] Corresponding
 zonally averaged wind profile, characterized by a small
   number of broad jets (three in this case).}
\end{figure}

Adopting this point of view, \citet{cho-etal-2003,
 cho-etal-2007} solved the equivalent-barotropic formulation of the
primitive equations on the rotating sphere.  These equations
are a one-layer version of the primitive equations and can formally be
obtained by integrating the primitive equations vertically.
The vertical integration leads to equations for the 
column-integrated horizontal velocity, layer
thickness, and mean temperature as a function of longitude, latitude,
and time.  The advantage of this approach is that, by reducing
the vertical resolution, high horizontal resolution can be
achieved, allowing numerical solutions to resolve small-scale
turbulent interactions that are difficult to capture in a full
3D model. These equations (and the formally related shallow-water 
equations) are
well-suited to the study of two-dimensional jets, vortices, and
horizontally propagating gravity and Rossby waves.  These simplified
equations have a venerable history of successful use in studying
dynamical processes in the planetary and atmospheric sciences
\citep[for a review see][]{vasavada-showman-2005}. On the other hand,
the full three-dimensional nature of the flow is not represented.
These equations exclude 3D processes such as baroclinic instability,
vortex tilting, and vertical wave propagation.

Rather than exciting the flow only with a day-night heating contrast,
\citet{cho-etal-2003,cho-etal-2007} also initialize their simulations
with small-scale turbulent stirring.  Their flows subsequently evolve
to states containing meandering polar vortices, breaking Rossby waves,
and turbulent mixing (Fig.~3).  One may question the relevance of
small-scale initialization for hot Jupiter atmospheres, since they are
intensely heated on large, hemispheric scales.  In considering this
question, however, one should remember that coupling between small and
large scales in nonlinear flow is a delicate, and still not completely
understood, issue.  As an example, consider once again Uranus: it is
dominantly forced by insolation on a hemispheric scale
and yet the rotationally constrained flow is clearly {\it not} a
simple reflection of this external forcing.  It is significant that
equivalent-barotropic simulations with small-scale initial turbulence
are able to reproduce the dominant features of Uranus' atmospheric
flow \citep{cho-etal-2007}.  Such simulations demonstrate that
regardless of the type and scale of forcing, rotation plays a
dominant role in shaping the large-scale flows on planets.

\begin{figure}[!ht]
\hspace{-1cm}
\begin{minipage}[t]{4cm}
\begin{center}
\includegraphics[width=7cm, angle=0]{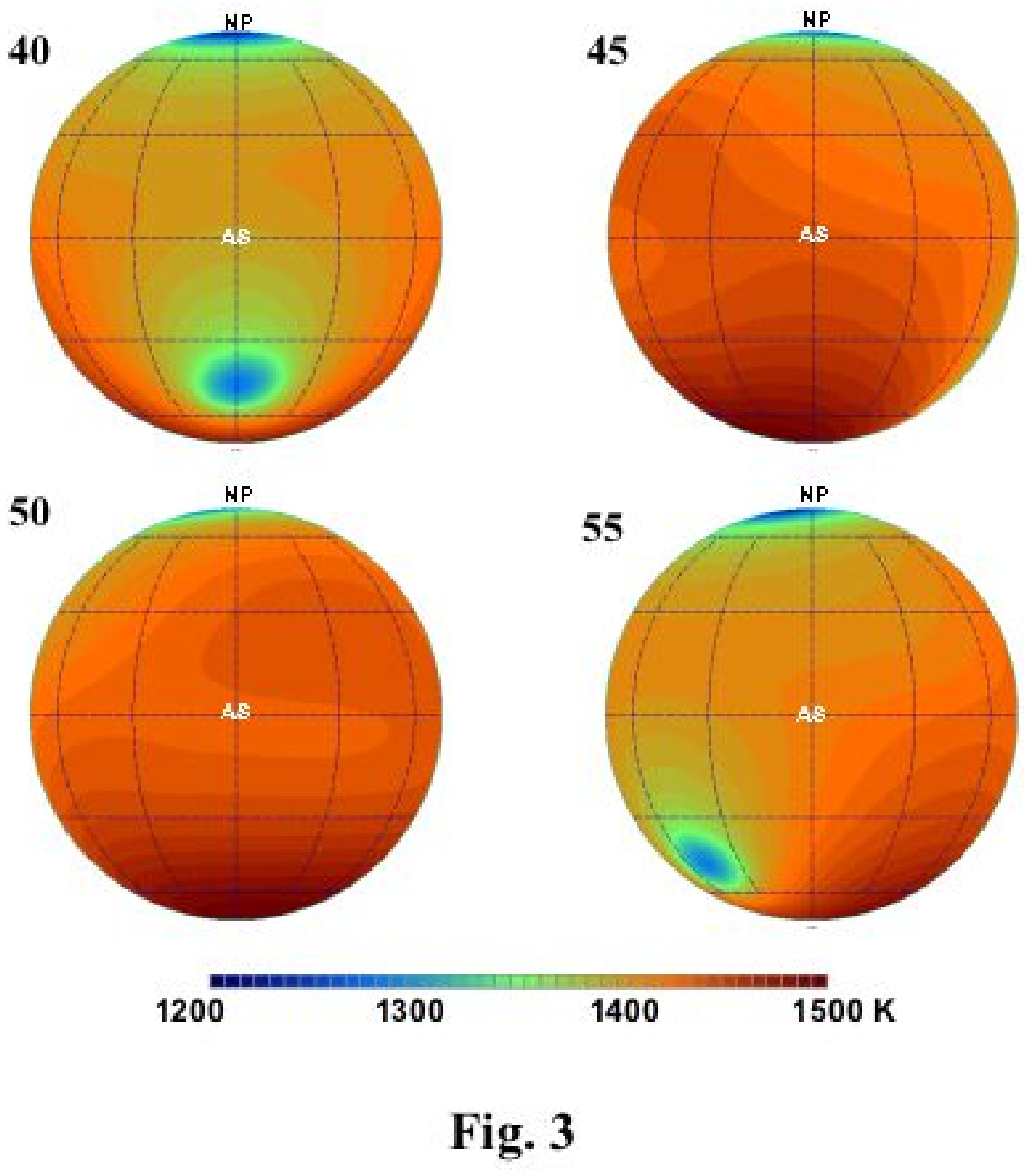}
\end{center}
\end{minipage}
\hspace{+2cm}
\begin{minipage}[t]{4cm}
\begin{center}
\includegraphics[width=7cm, angle=0]{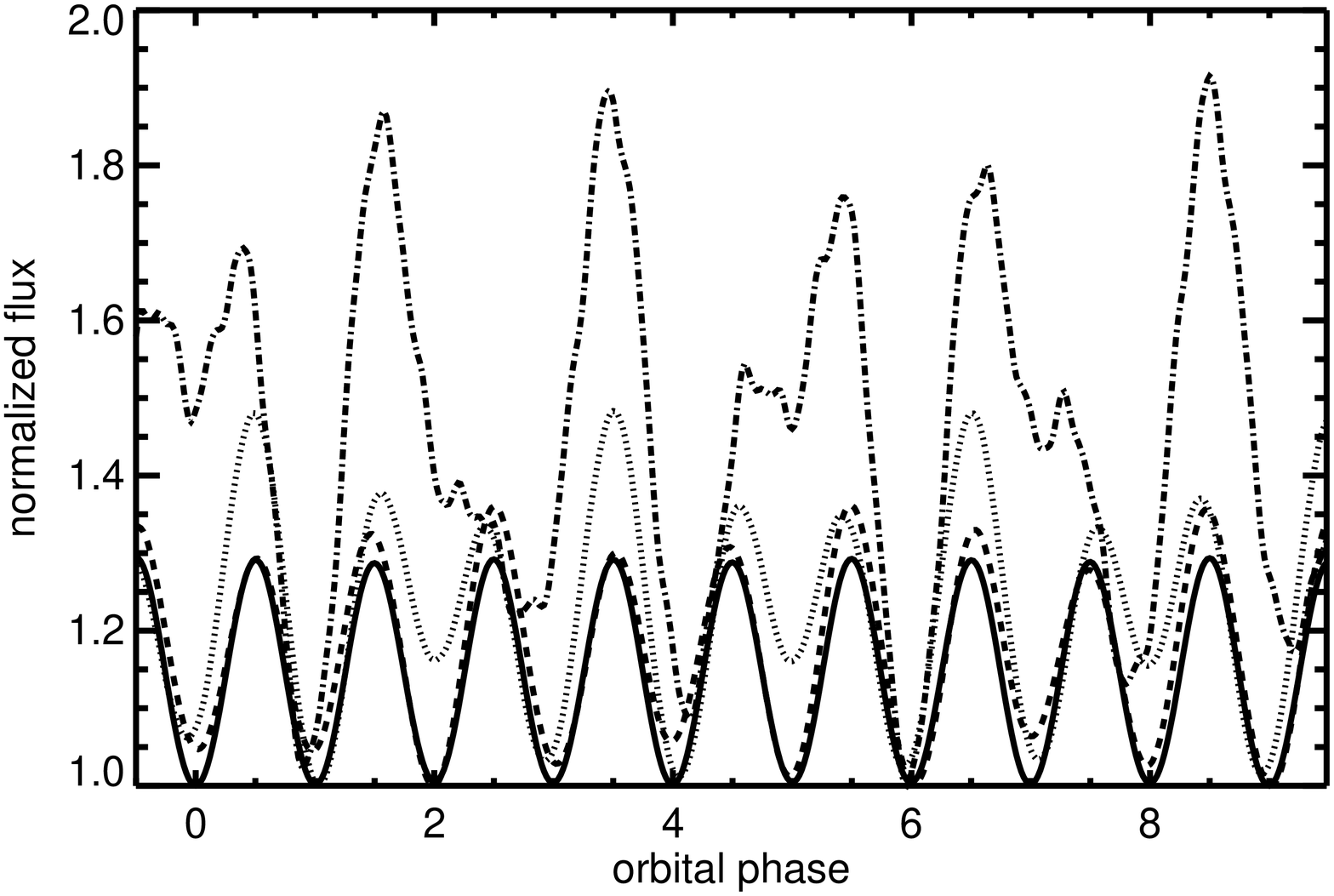}
\end{center}
\end{minipage}
\caption{[Left] Equatorial views of four successive temperature maps
for the same model as shown in Fig.~3. Large-scale moving atmospheric
features, in particular circumpolar vortices, may lead to significant
observable variability. [Right] Examples of thermal infrared phase
curves according to the models of \citet{cho-etal-2007}. From top to
bottom, curves for adopted mean wind speeds of $800$, $400$, $200$ and $100$
m/s are shown. At higher wind speeds, contributions from coherent
moving atmospheric features become apparent, shifting and deforming the
phase curves away from a regular shape \citep{rauscher-etal-2008}.}
\end{figure}

It is important to realize that the flow is stirred only at
 the start of the simulations in \citet{cho-etal-2003,cho-etal-2007}.
The resulting flow in these dynamically initialized studies can
 be extremely time-variable, especially when the mean winds
in the turbulent initial condition are strong.
 If such a behavior occurs on a hot Jupiter, then substantial
variability in the secondary eclipse depths of a transiting hot
 Jupiter may be expected  \citep{rauscher-etal-2007b}.  In addition,
  signatures of large-scale moving atmospheric features may become
 apparent (Fig.~4) in the thermal infrared phase curves of these
 objects \citep{rauscher-etal-2008}. In agreement with Rhines-length
and deformation-radius arguments \citep{showman-guillot-2002,
 cho-etal-2003, menou-etal-2003} and numerical simulations
by \citet{showman-guillot-2002} and \citet{cooper-showman-2005,
 cooper-showman-2006}, the flow usually consists of a small number of
broad jets (Fig.~3).

\subsection{Other work}

\citet{langton-laughlin-2007} performed global numerical simulations
of the atmospheric flow on hot Jupiters using the traditional
shallow-water equations, which govern the dynamics of a thin layer of
hydrostatically balanced, constant-density fluid on a rotating sphere.
Mathematically, these equations can be viewed as a special case of the
equivalent-barotropic equations in isentropic coordinates when
$R/c_p = 1$, where $R$ is the gas constant and $c_p$ is the
specific heat at constant pressure \citep{cho-etal-2007}.  However,
because the density in the shallow-water equations is constant
(whereas it is dependent on pressure in the equivalent barotropic
equations), the temperature is undefined and there is no thermodynamic
energy equation \citep[see][Chapter 3]{pedlosky-1987}.  Note that Eq.~1 of
\citet{langton-laughlin-2007}, which they write as an energy equation,
is actually the shallow-water mass continuity equation for the layer
thickness:
\begin{equation}
{\partial h\over \partial t} + \nabla\cdot (h {\bf v}) = 0,
\end{equation}
where $h$ is layer thickness (a representation of pressure), $t$ is
time, and ${\bf v}$ is horizontal velocity of the shallow-water layer.
\citet{langton-laughlin-2007} force Eq.~1 by including a
Newtonian-relaxation source term $(h_{\rm eq} - h)/\tau_{\rm rad}$,
where $h_{\rm eq}$ is the equilibrium thickness (thick on dayside,
thin on nightside), which adds mass on the dayside and removes it on
the nightside.  This is a reasonable zeroth-order method to 
model the effects of radiative cooling on the shallow-water flow
\citep[e.g.,][]{polvani-etal-1995, showman-2007}.  A rigorous
representation of thermodynamic forcing and temperature requires the
use of a more general set of equations (such as the
equivalent-barotropic, primitive, or compressible Navier-Stokes
equations) that, unlike the shallow-water equations, have distinct energy
equations.  In the planetary context, the shallow-water layer could be
interpreted as a thin, stable ``weather layer'', possibly overlying a
deeper and denser interior.
\citet{langton-laughlin-2007} adopt a radiative time constant of
$\sim$8\,hours, intended to represent conditions expected near
the photosphere according to \citet{iro-etal-2005}.
These forced simulations are initialized from
rest.\footnote{\citet{langton-laughlin-2007} also present unforced
  simulations initialized with small-scale turbulence, similar to the
  simulations of \citet{cho-etal-2003}.  In this case, they obtain a
  flow similar to \citet{cho-etal-2003}.}

\citet{langton-laughlin-2007} find that, when the obliquity is assumed
zero, their forced flows quickly reach a steady state exhibiting a
westward equatorial jet with speeds approaching $1\,{\rm km}\,{\rm
  sec}^{-1}$ and day-night differences in layer thickness reaching the
imposed equilibrium value of $\sim40\%$.  The height field becomes
distorted from the equilibrium value, $h_{\rm eq}$, so that the
thickest and thinnest regions are offset from the substellar and
antistellar points.  Interestingly, despite the development of a
westward equatorial jet, these thin and thick regions are displaced to
the {\it east} rather than to the west of the antistellar and 
substellar points, respectively; the reasons for this
phenomenon remain unclear.  \citet{langton-laughlin-2007} also
performed simulations with a high obliquity, wherein the starlight
alternately heats one hemisphere and then the other; the dynamics here
becomes periodic with the orbital period, as expected.  In all these
simulations, the flow length scales are broad, similar to the previous
studies discussed earlier.

Several authors have also performed investigations in limited-area
domains.  \citet{burkert-etal-2005} perform a 2D study (with
coordinates longitude and height) using the Navier-Stokes equations
without rotation in a section of an equatorial plane.  They
parameterize the heating with a flux-limited diffusion scheme using
Rosseland mean opacities and an imposed temperature distribution at
the top of the model typically ranging from $100\,$K on the nightside
to $\sim$1200\,K at the substellar point.  This scheme induces heating
on the dayside and cooling on the nightside.  Their standard domain
extends $180^{\circ}$ in longitude and $\sim$6000$\,{\rm km}$
vertically with free-slip wall boundary conditions.  Because of the
side wall boundary conditions, the flow cannot develop
planet-encircling jet streams \citep[as occurs for example in the
simulations of][] {showman-guillot-2002, cooper-showman-2005,
  cho-etal-2003, cho-etal-2007, dobbs-dixon-lin-2007}.  Instead, an
overturning circulation develops with rising motion near one wall,
sinking motion near the other, and lateral currents that connect the
two.  The nominal opacity used for calculating the diffusion
coefficients in the radiative scheme is based on that expected for
grains in the interstellar medium, although simulations were also
performed with opacities ranging from $10^{-3}$ to $10^2$ times the
nominal values.  The strength of the flow depends on the assumed
opacity; in the nominal cases, the wind speeds reach several ${\rm
  km}\,{\rm sec}^{-1}$ and day-night temperature differences reach
$\sim$700\,K at the expected photosphere pressure.

\citet{dobbs-dixon-lin-2007} investigated the atmospheric circulation
by solving the 3D Navier-Stokes equations in the low- and
mid-latitudes of a rotating sphere (the polar regions were omitted).
As in \citet{burkert-etal-2005}, they parameterized the
radiative-transfer with flux-limited diffusion using Rosseland-mean
opacities.  In addition to the Coriolis force, 
they included the centrifugal force in the equation
of motion; this differs from the approach taken in most 
global atmospheric circulation models, which typically
account for the gravitational relaxation of the planetary interior
to the planetary rotation by absorbing the centrifugal force
into the gravity \citep[see][p.~12--13]{holton-2004}.  Similar to
\citet{showman-guillot-2002} and \citet{cooper-showman-2005,
  cooper-showman-2006}, their simulations develop an eastward jet at
the equator and westward jets in midlatitudes.  The vertical thickness
of these jets varies substantially with longitude.
\citet{dobbs-dixon-lin-2007} find that the flow strength depends on
the assumed opacity, as in \citet{burkert-etal-2005}; in their nominal
cases, wind speeds reach several ${\rm km}\,{\rm sec}^{-1}$ and
day-night temperature differences at the expected photosphere reach
$\sim700$--$800\,$K.

The studies described so far all focus on flow near the photosphere.
In contrast, \citet{koskinen-etal-2007} performed 3D simulations of
the thermosphere and ionosphere of hot Jupiters using the 3D primitive
equations.  In their simulations, the domain extended from
$\sim2\,\mu$bar to $\sim0.04\,$nbar or less.  The simulations included
molecular heat conduction and diffusion, which are important due to
the large molecular mean-free paths at the sub-$\mu$bar pressures
investigated in this study.  Simplified schemes for photochemistry,
stellar extreme ultraviolet (EUV) heating, and radiative cooling by
H$_3^+$ were also implemented.  The emphasis was on planets from
0.2--$1\,$AU from their stars (in contrast to the other investigations
described in this review, which tend to emphasize planets at
$<0.1\,$AU).  The simulations developed strong flows of $\sim$1$\,{\rm
  km}\,{\rm sec}^{-1}$ from the substellar point to the antistellar
point; temperatures often reached $10,000\,$K or more depending on
assumptions about the cooling.  The wind, temperature, and chemistry
patterns have implications for the rate of atmospheric escape to
space.

\subsection{Speculations on the causes of model differences}

Published simulations of atmospheric circulation on hot Jupiters have
similarities as well as differences.  Despite the range of approaches,
all the published studies agree that the flow on typical hot Jupiters
will exhibit only a small number of broad jets, consistent with Rhines
scale and deformation radius arguments \citep{showman-guillot-2002,
  cooper-showman-2005, cooper-showman-2006, menou-etal-2003,
  cho-etal-2003, cho-etal-2007, langton-laughlin-2007,
  dobbs-dixon-lin-2007}.  As discussed earlier, this agreement results
from the modest rotation rates, high temperatures, and strong static
stabilities of hot-Jupiter atmospheres.  (Indeed, this bodes well for
characterizing the dynamical state with spectra and lightcurves, which
sample globally or near-globally.)  Published simulations that attempt
to predict the wind speeds also generally agree that the baroclinic
component of the flow (i.e., the difference in the flow speeds between
the photosphere and the base of the radiative zone) may reach
$\sim1\,{\rm km}\,{\rm sec}^{-1}$ or more \citep{showman-guillot-2002,
  cooper-showman-2005, cooper-showman-2006, burkert-etal-2005,
  langton-laughlin-2007, dobbs-dixon-lin-2007}.

On the other hand, the models also exhibit a number of important
differences.  Characteristic features of the \citet{cho-etal-2003,
  cho-etal-2007} simulations include development of large polar
vortices, extensive turbulent mixing, and time variability, if the
global average wind speed is large and/or the day-night hemispheric
temperature difference is weak.  In contrast, the simulations of
\citet{showman-guillot-2002}, \citet{cooper-showman-2005,
  cooper-showman-2006}, and \citet{langton-laughlin-2007} exhibit
steadier flow behavior and tend to lack the high degree of turbulent
mixing that occurs in the \citeauthor{cho-etal-2007} simulations.
While differences in model resolution and numerical methods surely
contribute, one cause of these differences seems to be the issue of
whether the initial flow contains turbulence
\citep[as in][]{cho-etal-2003, cho-etal-2007}, in addition to a global-scale
day-night forcing \citep[as in][]{showman-guillot-2002,
  cooper-showman-2005, cooper-showman-2006, langton-laughlin-2007,
  dobbs-dixon-lin-2007}.  Certainly, if the initial condition contains
strong, closely-packed vortices (as in \citeauthor{cho-etal-2007}),
the subsequent process of vortex mergers and planetary wave radiation
naturally induces turbulent mixing and produces, as an end state, a
small number of large vortices that typically end up near the poles
\citep[see][] {cho-polvani-1996b}.
On the other hand, when the energy source
comprises global-scale day-night forcing, the dominant length scales are
large ($\sim$planetary radius), leading to a global-scale flow.
\footnote{Large-scale jet streams can exhibit shear
instabilities that generate small-scale turbulence, but this
phenomenon has not been obviously apparent in the radiatively forced 
studies performed to date.}

Therefore, a key distinction between the various models
published so far has to do with the problem of understanding how
the stellar forcing is expressed in the atmospheric flow itself.
As we have already commented, this is a subtle issue, even for
solar-system giant planets. However, there are good reasons to believe
that progress on this issue is possible in the future. First, on the
modeling side, detailed schemes to treat the radiative transfer in
hot Jupiter atmospheres are being developed, so that more reliable
treatments of the non-linear radiation-hydrodynamics involved in
expressing this forcing can be implemented. These modeling efforts
may thus be able to shed some light as to what the nature of the
flow forcing in hot Jupiter atmospheres is. Secondly, observations
may provide important guidance for the modelers, as is the case even
in solar-system atmospheric studies.\footnote{Even advanced GCMs,
which are treated as much as possible as first-principle models,
generally include numerous free parameters which are adjusted to
match observed atmospheres.} This is particularly true of the issue
of atmospheric turbulence.  All solar-system atmospheres are turbulent.
For hot Jupiters, there is an expectation that the combination of
turbulence and large fundamental atmospheric scales may lead to
significant variability in various observables. Observations may thus
 help distinguish a rather steady atmospheric behavior from a more
variable one, and in doing so establish a distinction between the
two modeling approaches (and associated forcing schemes) discussed
above. As has always been the case with solar-system studies, progress
in understanding hot Jupiters is likely to involve repeated iterations
between observations and modeling.

Another difference in published studies is the development of robust
equatorial superrotation in the 3D simulations
\citep{showman-guillot-2002, cooper-showman-2005, cooper-showman-2006,
  dobbs-dixon-lin-2007}, whereas the one-layer calculations produce
both westward (i.e., subrotating) and eastward equatorial flow
\citep{cho-etal-2003, cho-etal-2007, langton-laughlin-2007}.
Regarding the superrotation, the studies by
\citet{showman-guillot-2002} and \citet{cooper-showman-2005,
  cooper-showman-2006} involved a range of different assumptions for
the initial temperature profile, the day-night equilibrium temperature
difference, and the radiative time constants.  Some simulations were
initialized with a strong {\it westward} jet to determine if it could
be retained in the simulation.  Intriguingly, all of these simulations
produced a strong eastward jet at the equator.  This suggests that the
superrotation is a robust phenomenon, at least within the context of
the adopted input parameters and forcing approach.  More study is
needed to understand this phenomenon.

The forcing in all published models is simplified; no models yet
include realistic representations of radiative transfer.
\citet{burkert-etal-2005} and \citet{dobbs-dixon-lin-2007} adopted a
diffusion approach, which is accurate in the deep, optically thick
atmosphere (and hence is commonly used in the interior portion of
evolution models); however, the radiative transfer is non-diffusive
above the photosphere, so this approach loses accuracy in the
observable atmosphere (i.e. at and above the photosphere).  In
contrast, the shallow-water studies of \citet{langton-laughlin-2007}
and the 3D studies of \citet{showman-guillot-2002} and
\citet{cooper-showman-2005, cooper-showman-2006} adopted a Newtonian
heating/cooling scheme.  This scheme has enjoyed a long history of
successful use in process studies of planetary climate.  However, it
effectively involves a linearization of the heating rate around the
radiative-equilibrium state and neglects the fact that the radiative
equilibrium temperature and timescale can depend on the atmosphere's
dynamical response, particularly when actual temperatures are far from
the prescribed radiative-equilibrium values.
Finally, \citet{cho-etal-2003, cho-etal-2007} did not include
radiative forcing at all but instead performed adiabatic simulations
driven by a combination of initial turbulence and large-scale
pressure deflections.  This approach allows an exploration of 
how the flow behavior depends on the energy (which is a tunable 
input parameter) and therefore provides a useful baseline.  
On the other hand, there is no ability to predict mean 
flow speeds, and the absence of radiative forcing
means that these simulations cannot capture aspects of the dynamics
that crucially depend on such forcing. The adopted approaches, while
simplified, nevertheless all serve the intended purpose of providing
plausible sources of energy injection.  Inclusion of realistic
radiative transfer is an important goal for future modeling
work.

Interestingly, infrared lightcurve observations
could allow an observational constraint on the wind patterns,
at least for planets with a photosphere deep
enough for the radiative time to be comparable to the advection
time.   To varying degrees, the simulations performed to date 
suggest that dynamics can distort the temperature pattern in
a complex manner, so that the temperature field does not 
simply track the stellar heating pattern \citep{showman-guillot-2002,
cooper-showman-2005, cooper-showman-2006, cho-etal-2003, cho-etal-2007,
burkert-etal-2005, dobbs-dixon-lin-2007, langton-laughlin-2007}.
For example, the hottest and coldest regions may be
displaced from the substellar and antistellar points, respectively,
leading to phase shifts in infrared lightcurves relative to
that expected without dynamics.  While there will likely be degeneracies
in interpretation (i.e., multiple flow patterns may explain
a given lightcurve), such data will nevertheless provide
powerful contraints to distinguish among the various models.

\subsection{The validity of hydrostatic balance}

A scaling analysis demonstrates that local hydrostatic balance is approximately
valid for the large-scale flow on hot Jupiters.  It is important to emphasize
that the local-hydrostatic-balance assumption in the primitive 
equations derives from the
assumption of large aspect ratio and {\it not} from
any assumption on wind speed.  However, the fact that estimated
wind speeds on hot Jupiters are several ${\rm km}\,{\rm sec}^{-1}$, which
is close to the 3-km$\,{\rm sec}^{-1}$ speed of sound in these
atmospheres, suggests that we consider the validity of the
primitive equations in the hot-Jupiter context. The full Navier-Stokes
vertical momentum equation can be written
\begin{equation}
{\partial w\over\partial t} + {\bf v}\cdot\nabla w =-{1\over\rho}
{\partial p\over\partial z} - g + 2 u \Omega \cos\phi
\label{vert-mom}
\end{equation}
where $w$ is vertical wind speed, $t$ is time, ${\bf v}$ is the
horizontal wind velocity, and $u$ is the east-west wind speed.  
The background static hydrostatic balance is irrelevant
to atmospheric circulation and can be removed from the equation.  Define
 $p=p_0(z)+p'$ and $\rho=\rho_0(z)+\rho'$ where $p_0$ and $\rho_0$
are the time-independent basic-state pressure and density and, 
by construction, ${\partial p_0/\partial z}\equiv -\rho_0 g$.  
Primed quantities
are the deviations from this basic state caused by dynamics.
Substituting these expressions into Eq.~\ref{vert-mom}, we can
rewrite the equation as
\begin{equation}
{\partial w\over\partial t} + {\bf v}\cdot\nabla w =-{1\over\rho}
\left({\partial p'\over\partial z} - \rho'g\right) + 2 u \Omega \cos\phi
\label{vert-mom2}
\end{equation}
where the basic-state hydrostatic balanced has been subtracted off.
The terms in parentheses give only the flow-induced contributions 
to vertical pressure gradient and weight 
(they go to zero in a static atmosphere).

For local hydrostatic balance to be a reasonable approximation, the
terms in parenthesis must be much larger than the other terms
in the equation.  The magnitude of $\partial w/\partial t$ is 
approximately $w/\tau$, where $\tau$ is a flow evolution timescale, 
and the magnitude of ${\bf v}\cdot\nabla w$ is the greater of
$uw/L$ and $w^2/H$ where $L$ is the horizontal flow scale and
$H$ is the vertical flow scale.  For global-scale flows,
$L\sim10^7$--$10^8\,$m and $\tau\sim10^4$--$10^5\,$sec.  
The large-scale
flow varies vertically over a scale height $H\sim300\,{\rm km}$ 
\citep{showman-guillot-2002, cooper-showman-2005, 
cooper-showman-2006, dobbs-dixon-lin-2007}.  
For the simulated flow regime in Cooper 
and Showman (2005, 2006) and Dobbs-Dixon and Lin (2007),
${\bf v} \sim u \sim 3\,{\rm km}\,{\rm sec}^{-1}$, $g\sim10\,{\rm m}
\,{\rm sec}^{-2}$, $\Omega\sim 2\times10^{-5}\,{\rm sec}^{-1}$, 
and $w\sim10$--$100\,{\rm m}\,{\rm sec}^{-1}$.
With these values, we find that $\partial w/\partial t
\le 10^{-2}\,{\rm m}\,{\rm sec}^{-2}$, $uw/L\le0.03\,{\rm m}\,{\rm sec}^{-2}$,
$w^2/H\le0.03\,{\rm m}\,{\rm sec}^{-2}$, and 
$\Omega u \sim 0.1\,{\rm m}\,{\rm sec}^{-2}$.
In comparison, for a hot Jupiter with day-night temperature
differences of several hundred K, the flow-induced hydrostatic terms 
$\rho'g/\rho$ and $\rho^{-1}\partial p'/\partial z$ are 
each $\sim10\,{\rm m}\,{\rm sec}^{-2}$.

This analysis implies that, for global-scale hot-Jupiter flows, 
the greatest departure from hydrostaticity results from the 
vertical Coriolis force,
which causes a $\sim1$\% deviation from hydrostatic balance.
The acceleration terms on the left side of Eq.~\ref{vert-mom2}
(which are necessary for vertically propagating sound
waves) cause a $\sim0.3\%$ deviation from hydrostatic balance.
Hydrostatic balance is thus a reasonable approximation for
the large-scale flow.  Nonhydrostatic effects of course become
important at small scales, and it is conceivable that these effects
interact with the large-scale flow in nontrivial ways.
For hot Jupiters, the acceleration terms on the left side
of Eq.~\ref{vert-mom2} only become important for
structures with vertical or horizontal scales less than
$\sim30\,$km and $\sim500$--$1000\,$km, respectively.  
In a numerical model that solved the
full Navier-Stokes equations, the grid resolution
would have to be substantially finer than these values for
the nonhydrostatic behavior to be accurately 
represented.  A full Navier-Stokes solution with a
coarse resolution would effectively be resolving just
the global-scale hydrostatic component of the flow.

\subsection{The need for model validation}

Due to the complexity of processes involved in modeling the coupled
radiation-hydrodynamics of a planetary atmosphere, it is customary in
the field of geophysical fluid dynamics and atmospheric sciences to
implement robust validation schemes for any simulation
tool. Typically, both radiative transfer modules and dynamical cores
(i.e., \textbf{``}primitive equation solvers\textbf{''}) are subjected
to standard test cases or inter-comparisons aimed at
quantifying their performance and accuracy in conserving
important quantities, such as energy and enstrophy (``vortical
  energy'').
For example, a classical benchmark test for dynamical cores is the
Held-Suarez test, which isolates baroclinic dynamics from
radiative transfer issues by using a simple, Newtonian
cooling scenario (Held and Suarez 1994).

The issue of model validation is a particularly critical one
for hot Jupiter studies.  Earth and planetary
scientists have access to detailed information on the atmospheres they
study, allowing serious model flaws to be rapidly identified.  This is
not the case with hot Jupiter atmospheres.  While information
available from observation of these distant atmospheres is rapidly
  increasing, it still remains subject to degeneracies in
interpretation.  This is so even if ``eclipse maps'' eventually
become available \citep{williams-etal-2006, rauscher-etal-2007a}.  As a
result, rigorous model validations using test cases and Solar
System examples will be critical.  Successful model validations 
can be achieved by explicitly reproducing known features in the 
Solar System planet atmospheres \citep[e.g.][]{cho-etal-2007}, 
using simulation tools which have been well-tested previously
\citep[e.g.][]{showman-guillot-2002,
cho-etal-2003,cooper-showman-2005} or by designing
appropriate inter-comparison tests tailored for hot
  Jupiter problems.

\section{Summary}

The study of hot Jupiter atmospheres is maturing.  While data of
increasing quality are being collected, atmospheric models are also
being refined to help build a robust, hierarchical understanding
of these unusual atmospheres.  Indeed, one of the main motivations
for studying these atmospheres, which is also a main source of
difficulty, is the unusual physical regime that characterizes
them.  Hence, hot Jupiters represent new laboratories for studying
the complex physics of giant planet atmospheres.  In this way,
they offer the promise of extending the boundary of comparative
planetology well beyond the solar-system planets.

\acknowledgements
We thank the editors for allowing us to write a joint review paper.
The preparation of this paper was supported by a NASA Planetary
Atmospheres grant to APS and NASA contract NNG06GF55G to KM.



\begin{thebibliography}{}


\bibitem[{{Baraffe} et~al.(2003){\it {Baraffe}, {Chabrier}, {Barman}, {Allard},
  and {Hauschildt}\/}}]{baraffe-etal-2003}
{Baraffe}, I., G.~{Chabrier}, T.~S. {Barman}, F.~{Allard}, and P.~H.
  {Hauschildt}, 2003:
\newblock {\it \aap\/}, {\bf 402}, 701--712.

\bibitem[{{Barman} et~al.(2005){\it {Barman}, {Hauschildt}, and
  {Allard}\/}}]{barman-etal-2005}
{Barman}, T.~S., P.~H. {Hauschildt}, and F.~{Allard}, 2005:
\newblock {\it \apj\/}, {\bf 632}, 1132--1139.

\bibitem[{{Bodenheimer} et~al.(2001){\it {Bodenheimer}, {Lin}, and
  {Mardling}\/}}]{bodenheimer-etal-2001}
{Bodenheimer}, P., D.~N.~C. {Lin}, and R.~A. {Mardling}, 2001:
\newblock {\it \apj\/}, {\bf 548}, 466--472.

\bibitem[{{Bodenheimer} et~al.(2003){\it {Bodenheimer}, {Laughlin}, and
  {Lin}\/}}]{bodenheimer-etal-2003}
{Bodenheimer}, P., G.~{Laughlin}, and D.~N.~C. {Lin}, 2003:
\newblock {\it \apj\/}, {\bf 592}, 555--563.

\bibitem[{{Burkert} et~al.(2005){\it {Burkert}, {Lin}, {Bodenheimer}, {Jones},
  and {Yorke}\/}}]{burkert-etal-2005}
{Burkert}, A., D.~N.~C. {Lin}, P.~H. {Bodenheimer}, C.~A. {Jones}, and H.~W.
  {Yorke}, 2005:
\newblock {\it \apj\/}, {\bf 618}, 512--523.

\bibitem[{{Burrows} et~al.(2005){\it {Burrows}, {Hubeny}, and
  {Sudarsky}\/}}]{burrows-etal-2005}
{Burrows}, A., I.~{Hubeny}, and D.~{Sudarsky}, 2005:
\newblock {\it \apjl\/}, {\bf 625}, L135--L138.

\bibitem[{{Chabrier} et~al.(2004){\it {Chabrier}, {Barman}, {Baraffe},
  {Allard}, and {Hauschildt}\/}}]{chabrier-etal-2004}
{Chabrier}, G., T.~{Barman}, I.~{Baraffe}, F.~{Allard}, and P.~H. {Hauschildt},
  2004:
\newblock {\it \apjl\/}, {\bf 603}, L53--L56.

\bibitem[{{Charbonneau} et~al.(2002){\it {Charbonneau}, {Brown}, {Noyes}, and
  {Gilliland}\/}}]{charbonneau-etal-2002}
{Charbonneau}, D., T.~M. {Brown}, R.~W. {Noyes}, and R.~L. {Gilliland}, 2002:
\newblock {\it \apj\/}, {\bf 568}, 377--384.

\bibitem[{{Charbonneau} et~al.(2007){\it {Charbonneau}, {Brown}, {Burrows}, and
  {Laughlin}\/}}]{charbonneau-etal-2007}
{Charbonneau}, D., T.~M. {Brown}, A.~{Burrows}, and G.~{Laughlin}, 2007:
\newblock  {\it Protostars and Planets V\/}, B.~{Reipurth}, D.~{Jewitt}, and
  K.~{Keil}, Eds., pp. 701--716.

\bibitem[{{Charbonneau} et~al.(2005){\it {Charbonneau},
  et~al.\/}}]{charbonneau-etal-2005}
{Charbonneau}, D., et~al., 2005:
\newblock {\it \apj\/}, {\bf 626}, 523--529.

\bibitem[{{Cho} and {Polvani}(1996{\natexlab{a}}){\it {Cho} and
  {Polvani}\/}}]{cho-polvani-1996a}
{Cho}, J.~Y.-K., and L.~M. {Polvani}, 1996{\natexlab{a}}:
\newblock {\it Science\/}, {\bf 8}(1), 1--12.

\bibitem[{{Cho} and {Polvani}(1996{\natexlab{b}}){\it {Cho} and
  {Polvani}\/}}]{cho-polvani-1996b}
{Cho}, J.~Y.-K., and L.~M. {Polvani}, 1996{\natexlab{b}}:
\newblock {\it Physics of Fluids\/}, {\bf 8}, 1531--1552.

\bibitem[{{Cho} et~al.(2001){\it {Cho}, {de la Torre Ju{\'a}rez}, {Ingersoll},
  and {Dritschel}\/}}]{cho-etal-2001}
{Cho}, J.~Y.-K., M.~{de la Torre Ju{\'a}rez}, A.~P. {Ingersoll}, and D.~G.
  {Dritschel}, 2001:
\newblock {\it J.\ Geophys.\ Res.\/}, {\bf 106}, 5099--5106.

\bibitem[{{Cho} et~al.(2003){\it {Cho}, {Menou}, {Hansen}, and
  {Seager}\/}}]{cho-etal-2003}
{Cho}, J.~Y.-K., K.~{Menou}, B.~M.~S. {Hansen}, and S.~{Seager}, 2003:
\newblock {\it \apjl\/}, {\bf 587}, L117--L120.

\bibitem[{{Cho} et~al.(2007){\it {Cho}, {Menou}, {Hansen}, and
  {Seager}\/}}]{cho-etal-2007}
{Cho}, J.~Y.-K., K.~{Menou}, B.~M.~S. {Hansen}, and S.~{Seager}, 2007:
\newblock {\it \apj\/}, submitted.
\newblock arXiv:astro-ph/0607338.

\bibitem[{{Cooper} and {Showman}(2005){\it {Cooper} and
  {Showman}\/}}]{cooper-showman-2005}
{Cooper}, C.~S., and A.~P. {Showman}, 2005:
\newblock {\it \apjl\/}, {\bf 629}, L45--L48.

\bibitem[{{Cooper} and {Showman}(2006){\it {Cooper} and
  {Showman}\/}}]{cooper-showman-2006}
{Cooper}, C.~S., and A.~P. {Showman}, 2006:
\newblock {\it \apj\/}, {\bf 649}, 1048--1063.

\bibitem[{{Cowan} et~al.(2007){\it {Cowan}, {Agol}, and
  {Charbonneau}\/}}]{cowan-etal-2007}
{Cowan}, N.~B., E.~{Agol}, and D.~{Charbonneau}, 2007:
\newblock {\it \mnras\/}, {\bf 379}, 641--646.

\bibitem[{{Deming} et~al.(2005){\it {Deming}, {Seager}, {Richardson}, and
  {Harrington}\/}}]{deming-etal-2005}
{Deming}, D., S.~{Seager}, L.~J. {Richardson}, and J.~{Harrington}, 2005:
\newblock {\it \nat\/}, {\bf 434}, 740--743.

\bibitem[{{Deming} et~al.(2006){\it {Deming}, {Harrington}, {Seager}, and
  {Richardson}\/}}]{deming-etal-2006}
{Deming}, D., J.~{Harrington}, S.~{Seager}, and L.~J. {Richardson}, 2006:
\newblock {\it \apj\/}, {\bf 644}, 560--564.

\bibitem[{{Dobbs-Dixon} and {Lin}(2007){\it {Dobbs-Dixon} and
  {Lin}\/}}]{dobbs-dixon-lin-2007}
{Dobbs-Dixon}, I., and D.~N.~C. {Lin}, 2007:
\newblock {\it \apj, {\rm submitted}\/}.

\bibitem[{{Fortney} et~al.(2005){\it {Fortney}, {Marley}, {Lodders}, {Saumon},
  and {Freedman}\/}}]{fortney-etal-2005}
{Fortney}, J.~J., M.~S. {Marley}, K.~{Lodders}, D.~{Saumon}, and R.~{Freedman},
  2005:
\newblock {\it \apjl\/}, {\bf 627}, L69--L72.

\bibitem[{{Fortney} et~al.(2006){\it {Fortney}, {Cooper}, {Showman}, {Marley},
  and {Freedman}\/}}]{fortney-etal-2006}
{Fortney}, J.~J., C.~S. {Cooper}, A.~P. {Showman}, M.~S. {Marley}, and R.~S.
  {Freedman}, 2006:
\newblock {\it \apj\/}, {\bf 652}, 746--757.

\bibitem[{{Friedson} and {Ingersoll}(1987){\it {Friedson} and
  {Ingersoll}\/}}]{friedson-ingersoll-1987}
{Friedson}, J., and A.~P. {Ingersoll}, 1987:
\newblock {\it Icarus\/}, {\bf 69}, 135--156.

\bibitem[{{Gierasch} et~al.(1997){\it {Gierasch},
  et~al.\/}}]{gierasch-etal-1997}
{Gierasch}, P.~J., et~al., 1997:
\newblock  {\it Venus II: Geology, Geophysics, Atmosphere, and Solar Wind
  Environment\/}, S.~W. {Bougher}, D.~M. {Hunten}, and R.~J. {Philips}, Eds.,
  pp. 459--500.

\bibitem[{{Guillot} and {Showman}(2002){\it {Guillot} and
  {Showman}\/}}]{guillot-showman-2002}
{Guillot}, T., and A.~P. {Showman}, 2002:
\newblock {\it \aap\/}, {\bf 385}, 156--165.

\bibitem[{{Guillot} et~al.(1996){\it {Guillot}, {Burrows}, {Hubbard}, {Lunine},
  and {Saumon}\/}}]{guillot-etal-1996}
{Guillot}, T., A.~{Burrows}, W.~B. {Hubbard}, J.~I. {Lunine}, and D.~{Saumon},
  1996:
\newblock {\it \apjl\/}, {\bf 459}, L35+.

\bibitem[{{Harrington} et~al.(2006){\it {Harrington}, {Hansen}, {Luszcz},
  {Seager}, {Deming}, {Menou}, {Cho}, and
  {Richardson}\/}}]{harrington-etal-2006}
{Harrington}, J., B.~M. {Hansen}, S.~H. {Luszcz}, S.~{Seager}, D.~{Deming},
  K.~{Menou}, J.~Y.-K. {Cho}, and L.~J. {Richardson}, 2006:
\newblock {\it Science\/}, {\bf 314}, 623--626.

\bibitem[{{Harrington} et~al.(2007){\it {Harrington}, {Luszcz}, {Seager},
  {Deming}, and {Richardson}\/}}]{harrington-etal-2007}
{Harrington}, J., S.~{Luszcz}, S.~{Seager}, D.~{Deming}, and L.~J.
  {Richardson}, 2007:
\newblock {\it \nat\/}, {\bf 447}, 691--693.

\bibitem[{{Held}(2005){\it {Held}\/}}]{held-2005}
{Held}, I., 2005:
\newblock {\it Bull. Amer. Meteorological Soc.\/}, {\bf 86}.

\bibitem[{{Held} and {Suarez}(1994){\it {Held} and
  {Suarez}\/}}]{held-suarez-1994}
{Held}, I.~M., and M.~J. {Suarez}, 1994:
\newblock {\it Bulletin of the American Meteorological Society, vol.~75, Issue
  10, pp.1825-1830\/}, {\bf 75}, 1825--1830.

\bibitem[{{Holton}(2004){\it {Holton}\/}}]{holton-2004}
{Holton}, J.~R., 2004:
\newblock {\it An Introduction to Dynamic Meteorology, 4th Ed.\/}.
\newblock Academic Press, San Diego.

\bibitem[{{Huang} and {Robinson}(1998){\it {Huang} and
  {Robinson}\/}}]{huang-robinson-1998}
{Huang}, H.-P., and W.~A. {Robinson}, 1998:
\newblock {\it Journal of Atmospheric Sciences\/}, {\bf 55}, 611--632.

\bibitem[{{Ingersoll}(1990){\it {Ingersoll}\/}}]{ingersoll-1990}
{Ingersoll}, A.~P., 1990:
\newblock {\it Science\/}, {\bf 248}, 308--315.

\bibitem[{{Ingersoll} and {Cuzzi}(1969){\it {Ingersoll} and
  {Cuzzi}\/}}]{ingersoll-cuzzi-1969}
{Ingersoll}, A.~P., and J.~N. {Cuzzi}, 1969:
\newblock {\it Journal of Atmospheric Sciences\/}, {\bf 26}, 981--985.

\bibitem[{{Ingersoll} and {Porco}(1978){\it {Ingersoll} and
  {Porco}\/}}]{ingersoll-porco-1978}
{Ingersoll}, A.~P., and C.~C. {Porco}, 1978:
\newblock {\it Icarus\/}, {\bf 35}, 27--43.

\bibitem[{{Iro} et~al.(2005){\it {Iro}, {B{\'e}zard}, and
  {Guillot}\/}}]{iro-etal-2005}
{Iro}, N., B.~{B{\'e}zard}, and T.~{Guillot}, 2005:
\newblock {\it \aap\/}, {\bf 436}, 719--727.

\bibitem[{{Juckes}(1989){\it {Juckes}\/}}]{juckes-1989}
{Juckes}, M., 1989:
\newblock {\it Journal of Atmospheric Sciences\/}, {\bf 46}, 2934--2956.

\bibitem[{{Juckes} and {McIntyre}(1987){\it {Juckes} and
  {McIntyre}\/}}]{juckes-mcintyre-1987}
{Juckes}, M.~N., and M.~E. {McIntyre}, 1987:
\newblock {\it \nat\/}, {\bf 328}, 590--596.

\bibitem[{{Knutson} et~al.(2007){\it {Knutson}, {Charbonneau}, {Allen},
  {Fortney}, {Agol}, {Cowan}, {Showman}, {Cooper}, and
  {Megeath}\/}}]{knutson-etal-2007b}
{Knutson}, H.~A., D.~{Charbonneau}, L.~E. {Allen}, J.~J. {Fortney}, E.~{Agol},
  N.~B. {Cowan}, A.~P. {Showman}, C.~S. {Cooper}, and S.~T. {Megeath}, 2007:
\newblock {\it \nat\/}, {\bf 447}, 183--186.

\bibitem[{{Koskinen} et~al.(2007){\it {Koskinen}, {Aylward}, {Smith}, and
  {Miller}\/}}]{koskinen-etal-2007}
{Koskinen}, T.~T., A.~D. {Aylward}, C.~G.~A. {Smith}, and S.~{Miller}, 2007:
\newblock {\it \apj\/}, {\bf 661}, 515--526.

\bibitem[{{Langton} and {Laughlin}(2007){\it {Langton} and
  {Laughlin}\/}}]{langton-laughlin-2007}
{Langton}, J., and G.~{Laughlin}, 2007:
\newblock {\it \apjl\/}, {\bf 657}, L113--L116.

\bibitem[{{Lian} and {Showman}(2007){\it {Lian} and
  {Showman}\/}}]{lian-showman-2007}
{Lian}, Y., and A.~P. {Showman}, 2007:
\newblock {\it Icarus, {\rm submitted}\/}.

\bibitem[{{Menou} et~al.(2003){\it {Menou}, {Cho}, {Seager}, and
  {Hansen}\/}}]{menou-etal-2003}
{Menou}, K., J.~Y.-K. {Cho}, S.~{Seager}, and B.~M.~S. {Hansen}, 2003:
\newblock {\it \apjl\/}, {\bf 587}, L113--L116.

\bibitem[{{Pedlosky}(1987){\it {Pedlosky}\/}}]{pedlosky-1987}
{Pedlosky}, J., 1987:
\newblock {\it Geophysical Fluid Dynamics, 2nd Ed.\/}.
\newblock Springer-Verlag, New York.

\bibitem[{{Peixoto} and {Oort}(1992){\it {Peixoto} and
  {Oort}\/}}]{peixoto-oort-1992}
{Peixoto}, J.~P., and A.~H. {Oort}, 1992:
\newblock {\it Physics of Climate\/}.
\newblock American Institute of Physics, New York.

\bibitem[{{Polvani} et~al.(1995){\it {Polvani}, {Waugh}, and
  {Plumb}\/}}]{polvani-etal-1995}
{Polvani}, L.~M., D.~W. {Waugh}, and R.~A. {Plumb}, 1995:
\newblock {\it Journal of Atmospheric Sciences\/}, {\bf 52}, 1288--1309.

\bibitem[{{Rauscher} et~al.(2007{\natexlab{a}}){\it {Rauscher}, {Menou}, {Cho},
  {Seager}, and {Hansen}\/}}]{rauscher-etal-2007a}
{Rauscher}, E., K.~{Menou}, J.~Y.-K. {Cho}, S.~{Seager}, and B.~M.~S. {Hansen},
  2007{\natexlab{a}}:
\newblock {\it \apjl\/}, {\bf 662}, L115--L118.

\bibitem[{{Rauscher} et~al.(2007{\natexlab{b}}){\it {Rauscher}, {Menou},
  {Seager}, {Deming}, {Cho}, and {Hansen}\/}}]{rauscher-etal-2007b}
{Rauscher}, E., K.~{Menou}, S.~{Seager}, D.~{Deming}, J.~Y.-K. {Cho}, and
  B.~M.~S. {Hansen}, 2007{\natexlab{b}}:
\newblock {\it \apj\/}, {\bf 664}, 1199--1209.

\bibitem[{{Rauscher} et~al.(2008){\it {Rauscher}, {Menou}, and
  {coauthors}\/}}]{rauscher-etal-2008}
{Rauscher}, E., K.~{Menou}, and {coauthors}, 2008:
\newblock {\it \apj, {\rm in preparation}\/}.

\bibitem[{{Read} and {Lewis}(2004){\it {Read} and {Lewis}\/}}]{read-lewis-2004}
{Read}, P.~L., and S.~R. {Lewis}, 2004:
\newblock {\it The Martian Climate Revisited\/}.
\newblock Springer/Praxis, New York.

\bibitem[{{Salby}(1989){\it {Salby}\/}}]{salby-1989}
{Salby}, M.~L., 1989:
\newblock {\it Tellus\/}, {\bf 41A}, 48.

\bibitem[{{Saravanan}(1993){\it {Saravanan}\/}}]{saravanan-1993}
{Saravanan}, R., 1993:
\newblock {\it Journal of Atmospheric Sciences\/}, {\bf 50}, 1211--1227.

\bibitem[{{Schneider}(2006){\it {Schneider}\/}}]{schneider-2006}
{Schneider}, T., 2006:
\newblock {\it Annual Review of Earth and Planetary Sciences\/}, {\bf 34},
  655--688.

\bibitem[{{Scott} and {Polvani}(2007){\it {Scott} and
  {Polvani}\/}}]{scott-polvani-2007}
{Scott}, R.~K., and L.~{Polvani}, 2007:
\newblock {\it J. Atmos. Sci\/}, {\bf 64}, 3158--3176.

\bibitem[{{Seager} et~al.(2005){\it {Seager}, {Richardson}, {Hansen}, {Menou},
  {Cho}, and {Deming}\/}}]{seager-etal-2005}
{Seager}, S., L.~J. {Richardson}, B.~M.~S. {Hansen}, K.~{Menou}, J.~Y.-K.
  {Cho}, and D.~{Deming}, 2005:
\newblock {\it \apj\/}, {\bf 632}, 1122--1131.

\bibitem[{{Showman}(2007){\it {Showman}\/}}]{showman-2007}
{Showman}, A.~P., 2007:
\newblock {\it J. Atmos. Sci.\/}, {\bf 64}, 3132--3157.

\bibitem[{{Showman} and {Guillot}(2002){\it {Showman} and
  {Guillot}\/}}]{showman-guillot-2002}
{Showman}, A.~P., and T.~{Guillot}, 2002:
\newblock {\it \aap\/}, {\bf 385}, 166--180.

\bibitem[{{Suarez} and {Duffy}(1992){\it {Suarez} and
  {Duffy}\/}}]{suarez-duffy-1992}
{Suarez}, M.~J., and D.~G. {Duffy}, 1992:
\newblock {\it Journal of Atmospheric Sciences\/}, {\bf 49}, 1541--1556.

\bibitem[{{Sukoriansky} et~al.(2007){\it {Sukoriansky}, {Dikovskaya}, and
  {Galperin}\/}}]{sukoriansky-etal-2007}
{Sukoriansky}, S., N.~{Dikovskaya}, and B.~{Galperin}, 2007:
\newblock {\it J. Atmos. Sci., in press\/}.

\bibitem[{{Vallis}(2006){\it {Vallis}\/}}]{vallis-2006}
{Vallis}, G.~K., 2006:
\newblock {\it Atmospheric and Oceanic Fluid Dynamics: Fundamentals and
  Large-Scale Circulation\/}.
\newblock Cambridge Univ. Press, Cambridge, UK.

\bibitem[{{Vallis} and {Maltrud}(1993){\it {Vallis} and
  {Maltrud}\/}}]{vallis-maltrud-1993}
{Vallis}, G.~K., and M.~E. {Maltrud}, 1993:
\newblock {\it J. Phys. Oceanography\/}, {\bf 23}, 1346--1362.

\bibitem[{{Vasavada} and {Showman}(2005){\it {Vasavada} and
  {Showman}\/}}]{vasavada-showman-2005}
{Vasavada}, A.~R., and A.~P. {Showman}, 2005:
\newblock {\it Reports of Progress in Physics\/}, {\bf 68}, 1935--1996.

\bibitem[{{Williams}(1978){\it {Williams}\/}}]{williams-1978}
{Williams}, G.~P., 1978:
\newblock {\it Journal of Atmospheric Sciences\/}, {\bf 35}, 1399--1426.

\bibitem[{{Williams}(1979){\it {Williams}\/}}]{williams-1979}
{Williams}, G.~P., 1979:
\newblock {\it Journal of Atmospheric Sciences\/}, {\bf 36}, 932--968.

\bibitem[{{Williams}(2003){\it {Williams}\/}}]{williams-2003a}
{Williams}, G.~P., 2003:
\newblock {\it Journal of Atmospheric Sciences\/}, {\bf 60}, 1270--1296.

\bibitem[{{Williams} et~al.(2006){\it {Williams}, {Charbonneau}, {Cooper},
  {Showman}, and {Fortney}\/}}]{williams-etal-2006}
{Williams}, P.~K.~G., D.~{Charbonneau}, C.~S. {Cooper}, A.~P. {Showman}, and
  J.~J. {Fortney}, 2006:
\newblock {\it \apj\/}, {\bf 649}, 1020--1027.

\bibitem[{{Yoden} and {Yamada}(1993){\it {Yoden} and
  {Yamada}\/}}]{yoden-yamada-1993}
{Yoden}, S., and M.~{Yamada}, 1993:
\newblock {\it Journal of Atmospheric Sciences\/}, {\bf 50}, 631--644.

\end{thebibliography}
\end{document}